\documentclass[aps,prb ,reprint,superscriptaddress]{revtex4-1}

\usepackage{graphicx}
\usepackage{epstopdf}
\usepackage{dcolumn}
\usepackage{bm}
\usepackage{amsmath}
\usepackage{hyperref}
\hypersetup{backref, colorlinks=true, linkcolor=blue, citecolor=blue, urlcolor=blue}
\usepackage{sistyle} 
\usepackage{multirow}

\draft

\begin{document}


\title {Structural dynamics during laser induced ultrafast demagnetization}


\author{Emmanuelle Jal}
\affiliation{Sorbonne Universités, UPMC Univ Paris 06, CNRS, Laboratoire de Chimie Physique - Matière et Rayonnement, 75005 Paris, France}

\author{Victor L\'opez-Flores}
\affiliation{Synchrotron SOLEIL, Saint-Aubin, Boite Postale 48, 91192 Gif-sur-Yvette Cedex, France}

\author{Niko Pontius}
\affiliation{Helmholtz-Zentrum Berlin f\"ur Materialien und Energie GmbH, Albert-Einstein-Stra{\ss}e 15, 12489 Berlin, Germany}

\author{Tom Fert\'e}
\affiliation{Universit\'e de Strasbourg, CNRS, Institut de Physique et Chimie des Mat\'eriaux de Strasbourg, UMR 7504, F-67000 Strasbourg, France}

\author{Nicolas Bergeard}
\affiliation{Universit\'e de Strasbourg, CNRS, Institut de Physique et Chimie des Mat\'eriaux de Strasbourg, UMR 7504, F-67000 Strasbourg, France}

\author{Christine Boeglin }
\affiliation{Universit\'e de Strasbourg, CNRS, Institut de Physique et Chimie des Mat\'eriaux de Strasbourg, UMR 7504, F-67000 Strasbourg, France}

\author{Boris Vodungbo}
\affiliation{Sorbonne Universités, UPMC Univ Paris 06, CNRS, Laboratoire de Chimie Physique - Matière et Rayonnement, 75005 Paris, France}

\author{Jan L\"uning}
\affiliation{Sorbonne Universités, UPMC Univ Paris 06, CNRS, Laboratoire de Chimie Physique - Matière et Rayonnement, 75005 Paris, France}
\affiliation{Synchrotron SOLEIL, Saint-Aubin, Boite Postale 48, 91192 Gif-sur-Yvette Cedex, France}

\author{Nicolas Jaouen}
\affiliation{Synchrotron SOLEIL, Saint-Aubin, Boite Postale 48, 91192 Gif-sur-Yvette Cedex, France}


\date{\today}

\begin{abstract}
The mechanism underlying femtosecond laser pulse induced ultrafast magnetization dynamics remains elusive despite two decades of intense research on this phenomenon. Most experiments focused so far on characterizing magnetization and charge carrier dynamics, while first direct measurements of structural dynamics during ultrafast demagnetization were reported only very recently. 
We here present our investigation of the infrared laser pulse induced ultrafast demagnetization process in a thin Ni film, which characterizes simultaneously magnetization and structural dynamics. This is achieved by employing femtosecond time resolved X-ray resonant magnetic reflectivity (tr-XRMR) as probe technique. The experimental results reveal unambiguously that the sub-picosecond magnetization quenching is accompanied by strong changes in non-magnetic X-ray reflectivity. These changes vary with reflection angle and changes up to 30$\%$ have been observed. Modeling the X-ray reflectivity of the investigated thin film, we can reproduce these changes by a variation of the apparent Ni layer thickness of up to 1$\%$. Extending these simulations to larger incidence angles we show that tr-XRMR can be employed to discriminate experimentally between currently discussed models describing the ultrafast demagnetization phenomenon.
\end{abstract}


\maketitle


\subsection{Introduction}
Controlling magnetization with light pulses has been a vision for decades that attracted the interest of a large scientific community with the discovery of the ultrafast demagnetization phenomenon in 1996.\cite{beaurepaire_ultrafast_1996} The realization of all-optical magnetization switching \cite{stanciu_all-optical_2007} about a decade later further stimulated world-wide intense experimental and theoretical research efforts. These activities gave rise to the development of several theoretical descriptions, but the scientific community remains undecided on the mechanism driving this phenomenon of significant technological relevance.\cite{ beaurepaire_ultrafast_1996, malinowski_control_2008, bigot_coherent_2009, koopmans_explaining_2010, battiato_superdiffusive_2010, boeglin_distinguishing_2010, radu_transient_2011, graves_nanoscale_2013, schellekens_investigating_2013, bergeard_ultrafast_2014} The key question, how the angular momentum of the ferromagnetic state is transferred out of the spin system on these ultrafast time scales, is therefore still unanswered. One of the most broadly discussed models is based on Elliott-Yafet-like electron-phonon spin-flip scattering.\cite{koopmans_explaining_2010} In this case, spin and lattice dynamics are strongly connected even on very short time scales. One may therefore expect new insight by probing experimentally not only magnetization and charge dynamics, but simultaneously lattice dynamics occurring on the sub- to few picosecond time scale of relevance for the ultrafast demagnetization process.

Since the advent of femtosecond pulsed X-ray sources, lattice dynamics have been investigated on ultrafast time scales in a variety of different scientific contexts.\cite{rose-petruck_picosecond-milliangstrom_1999, Fritz_ultrafast_2007, PhysRevB.78.060404, von_reppert_persistent_2016} A first study of lattice dynamics accompanying ultrafast demagnetization was recently realized by Henighan \textit{et al.}~\cite{henighan_generation_2016} on a thin Fe film grown epitaxially on MgO. These authors observed THz frequency oscillations of the scattering intensities, which they attribute to the excitation of coherent longitudinal acoustic phonons building up the strain wave generated by the infrared (IR) excitation pulse. This interpretation is supported by a time resolved electron diffraction study realized recently by Reid \textit{et al.}~\cite{reid_ultrafast_2016} on a free-standing film of FePt nanoparticles. The observed lattice parameter oscillations match once again the time scale of acoustic strain waves propagating forth and back through the thin film. 

In the above discussed studies different techniques were employed to investigate in two independent experiments magnetization and structural dynamics. To exclude experimental artifacts, e.g., due to limited reproducibility of experimental conditions, and to obtain even more detailed insight into the interplay between ultrafast magnetization and lattice dynamics, it is desirable to probe both dynamics simultaneously within a single experiment. One experimental technique exhibiting excellent sensitivity to magnetization and structural properties is X-ray resonant magnetic reflectivity (XRMR). While (non-resonant) X-ray reflectivity retrieves the charge density profile perpendicular to the sample surface, from which layer thickness and interface roughness can be retrieved, the magnetization depth profile becomes accessible when tuning the photon energy to a magnetically dichroic absorption resonance.\cite  {tonnerre_direct_2011, kortright_resonant_2013, jal_noncollinearity_2015,jal_magnetization_2013} Employing XRMR as a probe technique in a time resolved IR pump - X-ray probe experiment, it is thus possible to retrieve in a single experiment both, ultrafast magnetization and structural dynamics. The general feasibility of such an experiment was recently demonstrated at BESSY with a time resolution of 70 ps.\cite{tsuyama_photoinduced_2016}

We report here on our femtosecond (fs) time resolved XRMR (tr-XRMR) simultaneous investigation of magnetization, charge and lattice dynamics induced by a fs IR laser pump pulse in a prototype ferromagnetic thin film made of Ni. We find that significant changes in non-magnetic X-ray reflectivity accompany the sub-picosecond magnetization dynamics, which we can model by variations of the Ni film thickness. Furthermore, by modeling the tr-XRMR signal for larger incidence angles than accessible in our experiment we demonstrate that this technique will allow to discriminate between different mechanisms proposed to govern ultrafast demagnetization dynamics. Our study thus paves the way for future tr-XRMR experiments at X-ray free electron laser facilities, which provide the photon flux and energy resolution necessary to access the required X-ray incidence angles.

\subsection{Femtosecond time resolved XRMR}
For our tr-XRMR study we used  a prototype ferromagnetic thin film similar to those treated theoretically.\cite {koopmans_explaining_2010, battiato_superdiffusive_2010} It consist of a 15 nm thin polycrystalline Ni film which was grown by DC sputtering on top of a 35 nm thin metallic Pd layer. This buffer layer was deposited directly on the naturally grown silicon oxide layer of a 500 um thick Si substrate. To prevent oxidation the Ni film was capped with a 3 nm thin Pd layer. Static MOKE measurements verified the expected in-plane magnetization of the Ni film.  As discussed in more detail in section C, the structural composition of this multilayer sample was characterized by static x-ray reflectivity measurements realized at the METROLOGY \cite{idir_metrology_2010} and SEXTANTS \cite{sacchi_sextants_2013} beamline of Synchrotron SOLEIL.

\begin{center}
\begin{figure}
 \includegraphics[width=0.4\textwidth]{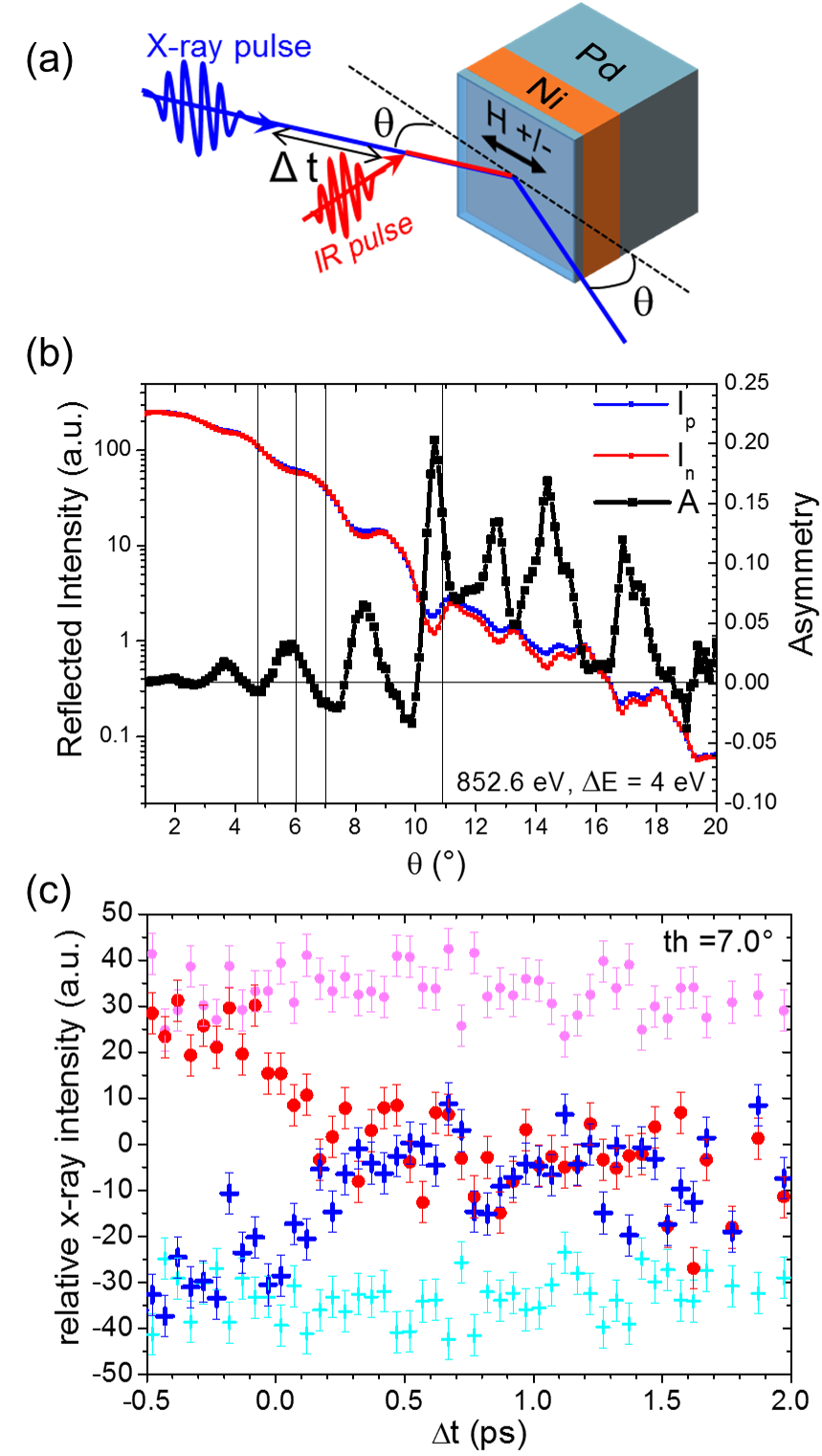}%
 \caption{(a) Sketch of the experimental geometry. (b) Static x-ray resonant magnetic reflectivity recorded at 852.6 $\pm2$ eV. The $I_p$ and $I_n$ reflected intensity for opposite in plane magnetization are shown by the blue and red curves respectively. The black curve gives the magnetic asymmetry. (c) Time resolved evolution of the reflected intensity as recorded for an incidence angle of $\theta=7.0\arcdeg$. Red circles and blue crosses give the intensities recorded for opposite in-plane magnetization directions with pumped and unpumped reference data plotted as full and light colors, respectively. The x-ray photon energy was 852.6 $\pm 2$ eV and the IR pump fluence 6 mJ/cm$^2$. \label{fig2}}
 \end{figure}
\end{center}

The tr-XRMR experiment was realized using the scattering end station of the FEMTOSPEX beamline at BESSY II.\cite{holldack_femtospex:_2014} As sketched in Fig.~\ref{fig2}(a) we used close to collinear in-coupling of the IR laser pulses  (800 nm, 50 fs FWHM) to avoid degradation of the time resolution due to different incidence angles of IR and X-ray pulses. The IR beam was focused to 0.40 mm x 0.25 mm such that the X-ray pulses (0.1 mm x 0.1 mm focus size) probe a rather homogeneously pumped sample area. The 6 kHz repetition rate of the slicing laser was split to record alternating pumped and unpumped reference data. For each pump-probe delay point, these signals were recorded for both in-plane magnetization directions ($I_p$ and $I_n$) and the halo background was subtracted as discussed in Ref.~\cite{schick_analysis_2016} As shown over the past years a time resolution of 130 fs is obtained routinely in the femtoslicing experiments,\cite{holldack_femtospex:_2014} which is predominantly set by the length of the sliced X-ray probe pulses. For all measurements the X-ray photon energy was set to match the magnetically dichroic Ni L$_3$ absorption edge (852.6 eV) with an energy resolution of E/$\Delta$E = 200, which we indicate as E $\pm2$ eV in the following.

Using the high photon flux in the picosecond mode of the FEMTOSPEX beamline \cite{holldack_femtospex:_2014} we aligned the experimental setup and recorded the static X-ray magnetic reflectivity curves $I_p$ and $I_n$ (blue and red curves in Fig.~\ref{fig2}(b)). The average of these two curves $I_{ave} = \frac{I_p + I_n}{2} = I_{NM}$ is sensitive to the (apparent) electronic charge density, \cite{mertins_resonant_2002, tonnerre_depth-resolved_2012} and is independent of the sample's magnetization. Indeed, the oscillations of $I_p$ and $I_n$ originate from interference of the x-rays reflected at the various interfaces of the sample. Therefore the analysis of the non-magnetic reflectivity, $I_{NM}$, allow us to extract structural parameters such as thickness, roughness and density of the different layers of our sample (see section C). The difference between the two curves $I_p$ and $I_n$ is however due to the sample magnetization. To reveal the magnetic term we derive the magnetic asymmetry as the difference of the two XRMR curves over their sum, $A= \frac{I_p - I_n}{I_p + I_n}$. This magnetic asymmetry, plotted by the black curve in Fig.~\ref{fig2}(b), is proportional to the ratio of magnetic to charge contribution.\cite{note} 

In the femtoslicing mode the accessible incidence angle range is limited by the available photon flux (about $10^{6}$ photons/sec). The choice of the incidence angles for the time resolved measurements is thus driven by the need of strong magnetic asymmetry and sufficient reflectivity to record XRMR curves within a reasonable data acquisition time. The vertical lines in Fig. \ref{fig2}(b) indicate the four chosen incidence angles $\theta$ = $4.8\arcdeg$, $6.0\arcdeg$, $7.0\arcdeg$ and $10.9\arcdeg$. For each of these angles we recorded the ultrafast evolution of the reflected intensity following IR laser excitation with a pump fluence of about 6 mJ/cm$^2$ for both opposite in-plane magnetization directions.

The ultrafast evolution of the reflected intensity is shown in the lower panel of Fig.~\ref{fig2} as function of the IR-pump - X-ray probe delay for the incidence angle of $\theta = 7.0\arcdeg$. The intensities recorded for opposite in-plane magnetization directions are shown by red circles and blue crosses. The IR laser pulse induced demagnetization manifests itself as the separation between pumped (full colors) and unpumped reference data (light colors). About 200 fs after the onset of this separation, the pumped data points recorded for the two opposite magnetization directions merge and remain so for the reminder of our time window. This implies that the magnetization is completely quenched within the probed subsurface layer of the film, an observation in line with previous results.\cite{koopmans_explaining_2010}

\begin{center}
\begin{figure}
 \includegraphics[width=0.4\textwidth]{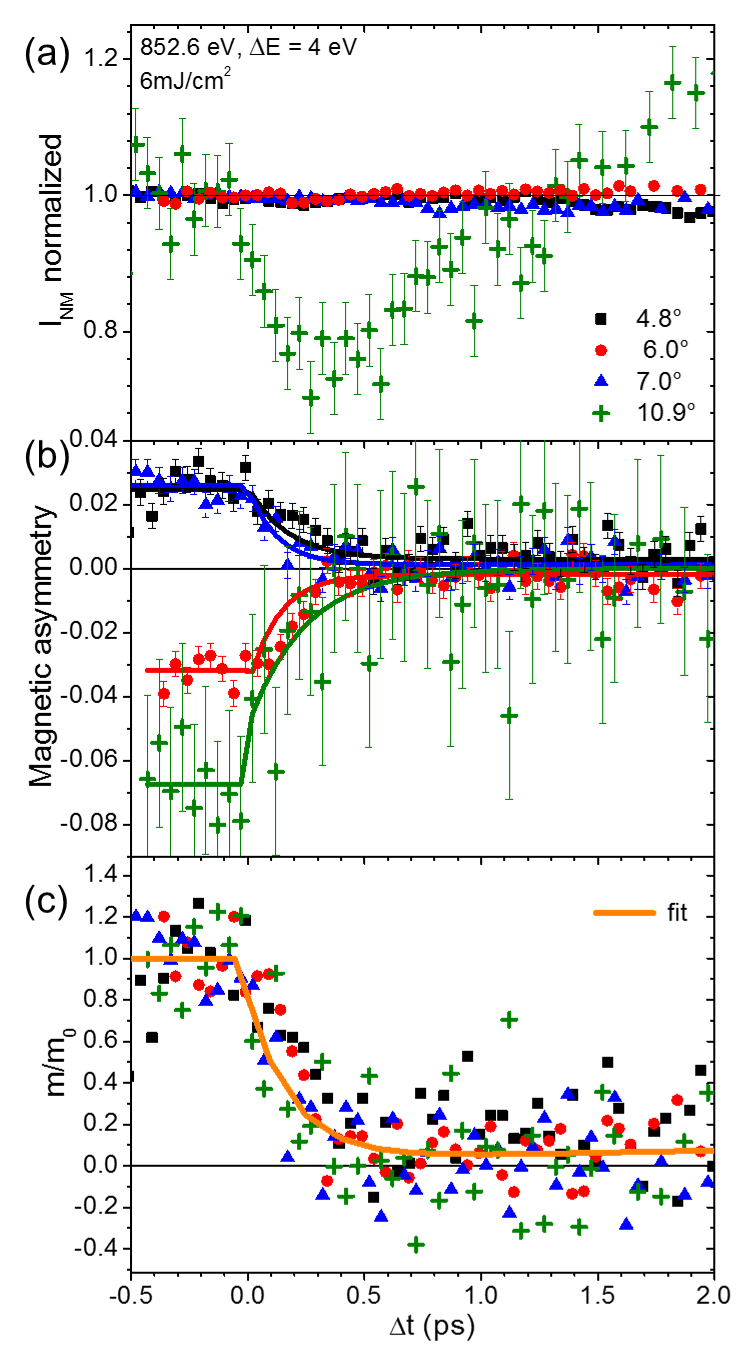}%
 \caption{(a) Evolution of the normalized non-magnetic reflectivity I$_{NM}$ with pump-probe delay as derived from the recorded reflected intensities for the four incidence angles of $\theta=4.8\arcdeg$ (black points), $\theta=6.0\arcdeg$ (red triangles), $\theta=7.0\arcdeg$ (blue diamonds) and $\theta=10.9\arcdeg$ (green crosses). (b) Evolution of the magnetic asymmetry with pump-probe delay as derived from the recorded reflected intensities for the four incidence angles. Lines indicate an exponential decay as a guide to the eye. (c) Relative change in magnetization derived from $I_{NM}$ and asymmetry curves in (a) and (b). All four magnetization dynamics can be fitted within the experimental accuracy by the same double exponential function, which is reproduced by the solid line. Experimental data were recorded with an x-ray photon energy of 852.6 $\pm 2$ eV and an IR pump fluence of 6 mJ/cm$^2$.
 \label{fig3}}
 \end{figure}
\end{center}

In order to compare our data for the four different incidence angles and to look separately at the charge and magnetic contribution we plot in the Fig.~\ref{fig3} the non-magnetic intensity $I_{NM}$ (a) and the magnetic asymmetry $A$ (b) as function of pump-probe delay. 

The dynamics of the non-magnetic contribution to the film's overall reflectivity is obtained by normalizing $I_{NM}$ to the unpumped reference data (Fig.~\ref{fig3}(a)). Surprisingly, the curve corresponding to an incidence angle of $\theta=10.9\arcdeg$ (green squares) exhibits strong changes. Note that sample degradation or drifts of the experimental setup can be ruled out as origin, since the in parallel recorded unpumped reference data remain unaltered. The amplitude of the variations of the non-magnetic reflectivity for the other incidence angles ($\theta=4.8\arcdeg$ (black points), $6.0\arcdeg$ (red triangles) and $7.0\arcdeg$ (blue diamonds)) are close to the noise limit; only the systematic deviation from the base line for longer delays may be significant.  

Looking at the time-resolved magnetic asymmetry in Fig.~\ref{fig3}(b), one notices that all four curves exhibit the usual shape of an ultrafast demagnetization process in the limit of strong pumping.\cite{koopmans_explaining_2010} From the magnetic asymmetry $A$ and the non-magnetic reflectivity $I_{NM}$ we can derive \cite{note} the purely magnetic contribution $m$, which is reproduced in Fig.~\ref{fig3}(c). The close similarity of these four curves suggests that the observed demagnetization dynamics does not depend within the given experimental accuracy on the incidence angle, i.e., on the thickness of the probed subsurface layer (the effective sampling depth varies from about 1.4 nm (4.8$\arcdeg$) to 4.0 nm (10.9$\arcdeg$), since the X-ray penetration length is about 20 nm at the Ni $L_3$ edge \cite{nakajima_electron-yield_1999}). 
Fitting these data with a common double exponential function \cite{malinowski_control_2008} yields a demagnetization time constant of $\tau_M$ = 170 $\pm$ 80 fs  (the recovery time $\tau_E$ was set to 17 ps \cite{koopmans_explaining_2010}) in agreement with previous results.\cite{koopmans_explaining_2010}

The strong changes of the non-magnetic reflectivity observed for an incidence angle of $\theta=10.9\arcdeg$ imply that the laser induced demagnetization goes along with simultaneously occurring changes of the geometric and/or electronic structure of the film. We remark that this finding is in line with previous direct observations of changes of the electronic structure \cite{stamm_femtosecond_2007, frietsch_disparate_2015} and the structural dynamics \cite{henighan_generation_2016, reid_ultrafast_2016} discussed in the introduction. 
In addition, we note that the absence of strong variations of the non-magnetic reflectivity observed for the three other incidence angles is not in contradiction to this conclusion, since reflectivity probes the film's properties differently at different incidence angles. Furthermore, since non-magnetic and magnetic X-ray reflectivity data can be quantitatively simulated,\cite{elzo_x-ray_2012} it is possible to test models predicting magnetization, structural and electronic dynamics. A detailed knowledge of the film's structural composition and its electronic and magnetic equilibrium properties is however needed. We therefore performed on the very same sample a static high resolution X-ray reflectivity study.

\begin{center}
\begin{figure}
 \includegraphics[width=0.45\textwidth]{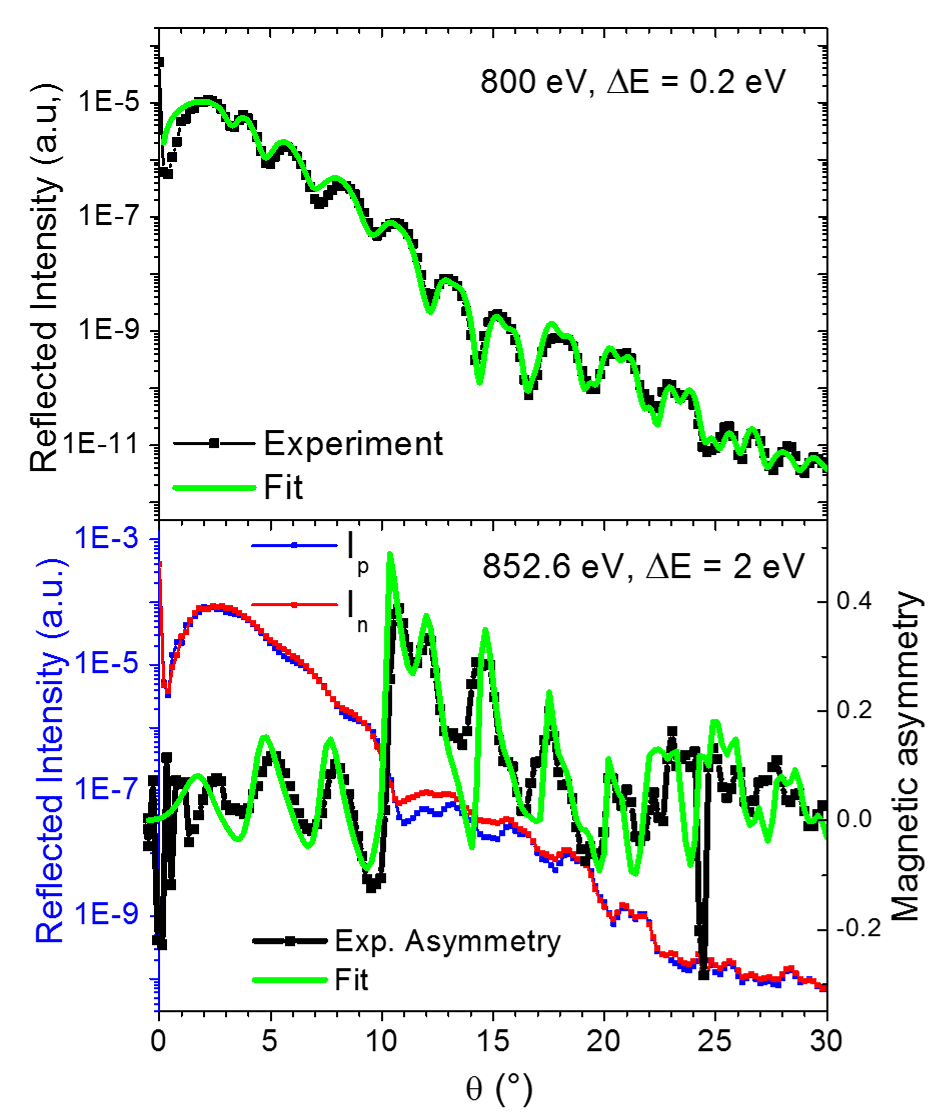}%
 \caption{(a) X-ray reflectivity curve (black square) recorded on the Ni film sample with circularly polarized X-rays of 800 eV (E/$\Delta$E = 4000). Modeling the film structure yields a simulation (green line) in excellent agreement with the experimental data. (b) X-ray reflectivity curves recorded with circular polarization at the magnetically dichroic L$_3$ edge of Ni (852.6 eV, E/$\Delta$E = 400) for the two opposite in-plane magnetization directions within the scattering plane (blue and red points). The derived asymmetry (black squares) is compared to its simulation (green line) based on the structural parameters derived from the data in (a). \label{fig1}}
 \end{figure}
\end{center}

\subsection{Static XRMR}
Non-resonant and resonant static X-ray reflectivity measurements were realized at the METROLOGY \cite{idir_metrology_2010} and SEXTANTS \cite{sacchi_sextants_2013} beamline of Synchrotron SOLEIL with a photon energy resolution of $\Delta$E = 0.2 eV and an angular resolution of 0.01$\arcdeg$. The end stations used for these measurements have been described elsewhere. \cite{jaouen_apparatus_2004, idir_metrology_2010} The black points in Fig.~\ref{fig1}(a) reproduce the non-resonant X-ray reflectivity curve of the sample recorded with circularly polarized X-rays of h$\nu$ = 800 eV, i.e., a photon energy well below the Ni L$_3$ resonance at 852 eV. Oscillations due to interference of the X-rays reflected at the various interfaces are clearly visible up to the highest measured reflection angle. This reveals the excellent flatness of the film layers and the high resolution/quality of the measurement. The main contributions to these oscillations can be identified by eye. The dominating periodicity of 2.4$\arcdeg$ is due to the 15 nm thin Ni layer. The superposed short period oscillation, particularly well visible at large incidence angles, is caused by the 35 nm thick buried   Pd buffer layer. And the 3nm Pd cap layer gives rise to the slow modulation barely visible having a period of about 14$\arcdeg$. To quantitatively analyze these data we have modeled the film's X-ray reflectivity with our implementation \cite{elzo_x-ray_2012} of the Paratt formalism.\cite{parratt_surface_1954} For the optical constants, we have used tabulated values for Pd and Si \cite{nist} and experimental values for Ni.\cite{chen_soft-x-ray_1990} The excellent agreement between simulation (green curve) and experimental data indicates that the structural composition of the film is well reproduced by the model. In table \ref{tab1} we list the fitted parameters characterizing this model. 

When using a photon energy in close vicinity to a magnetically dichroic X-ray absorption resonance, the optical constants depend also on the material's magnetization. It is thus possible to characterize within the same measurement the structural composition and the magnetization depth profile perpendicular to the film surface. This has been exploited before, e.g., to characterize orientation and magnitude of magnetic moments with element, site, and depth sensitivity.\cite{kortright_resonant_2013, tonnerre_depth-resolved_2012} Using circularly polarized X-rays with a photon energy matching the magnetically dichroic Ni L$_3$ absorption resonance at 852.6 eV, we have recorded the two reflectivity curves corresponding to the two opposite in-plane magnetization directions within the scattering plane (see Fig.~\ref{fig2}(a)). Note that these measurements were realized with a poor photon energy resolution of E/$\Delta$E = 400 to approach the experimental conditions of the FEMTOSPEX slicing facility. To visualize the magnetic contribution, we show in Fig.~\ref{fig1}(b) as black dots the derived magnetic asymmetry.  Note that the magnetic asymmetry is overall bigger than the one derived from the picosecond mode of the FEMTOSPEX beamline (Fig. \ref{fig2}(b)), since the photon resolving power at SOLEIL is two times higher than the one at FEMTOSPEX. 

Using the matrix formalism developed by Elzo \textit{et. al.} \cite{elzo_x-ray_2012} we can calculate the magnetic asymmetry expected for the above derived structural parameters using the known magnetically dichroic optical constants of Ni.\cite{chen_soft-x-ray_1990} The excellent agreement of this simulation (green line in Fig.~\ref{fig1}(b)) with the experimental data confirms the accuracy of our model of the structural composition and reveals an homogenous magnetization within the Ni film.

\newcolumntype{C}[1]{>{\centering\let\newline\\\arraybackslash\hspace{0pt}}m{#1}}

\begin{center}
\begin{table}
\begin{center}
\renewcommand{\arraystretch}{1}
\begin{tabular}{C{1cm} || C{2cm} | C{2cm} | C{2cm} }
                        & Density (mol.cm$^{-3}$) & Thickness (nm)      & Roughness (nm)   \\
\hline \hline Pd        & 0.115  $\pm0.002$       & 3.37 $\pm0.21$      & 0.69 $\pm0.06$   \\ 
\hline        Ni        & 0.155  $\pm0.003$       & 14.15 $\pm0.21$     & 0.70 $\pm0.02$   \\  
\hline        Pd        & 0.113                   & 35.30 $\pm0.5$      & 0.61 $\pm0.01$   \\   	
\hline        Si        & 0.083                   &                     & 0.49 $\pm0.09$   \\  
\end{tabular}
\caption{Structural parameters of the Ni film sample obtained by fitting the modeled non-resonant X-ray reflectivity to the data shown in Fig.~\protect\ref{fig1}(a). \label{tab1} }
\end{center}
\end{table}
\end{center}

\subsection{Structural dynamics}
\begin{center}
\begin{figure}
 \includegraphics[width=0.45\textwidth]{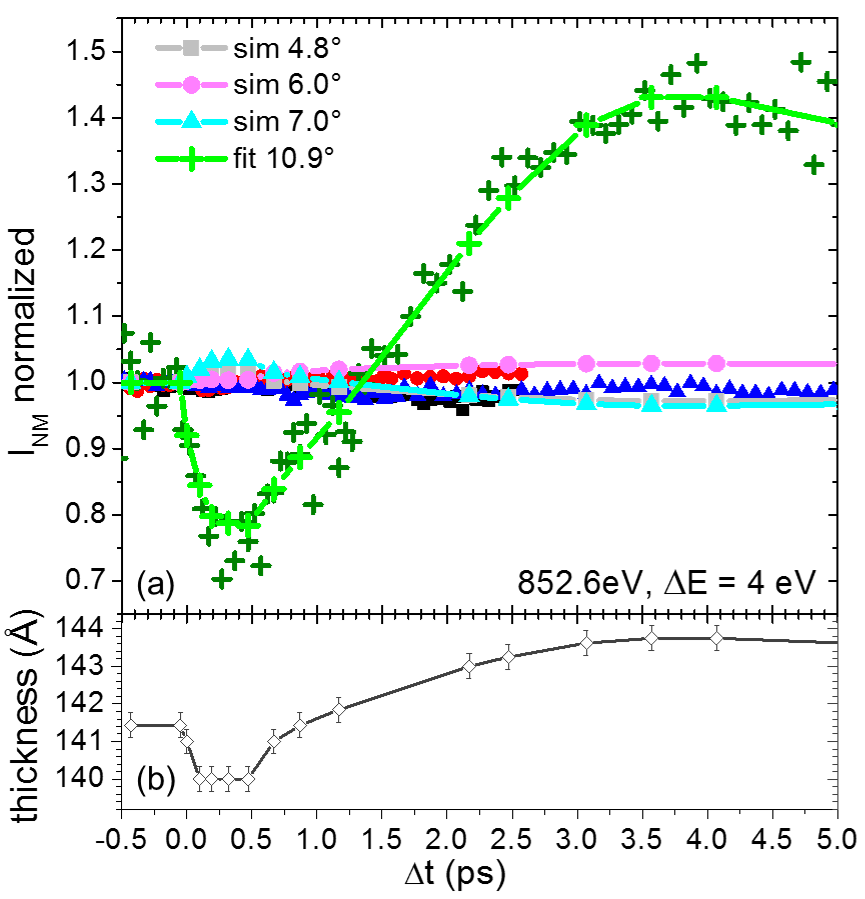}%
 \caption{(a) Comparison between measured (dark symbols taken from Fig.~\protect\ref{fig3}(c) for an extended time delay) and simulated (light symbols and connecting lines) temporal evolution of the non-magnetic reflectivity (I$_{NM}$) for an incidence angle of $\theta=4.8\arcdeg$ (black), $\theta=6.0\arcdeg$ (red), $\theta=7.0\arcdeg$ (blue) and $\theta=10.9\arcdeg$ (green). As discussed in detail in the text, all four simulations are realized with the same model, which assumes the temporal evolution of the Ni film thickness shown in panel (b). \label{fig4}}
 \end{figure}
\end{center}

Having such a detailed characterization at hand, we can now attempt to simulate the temporal evolution of the non-magnetic reflectivity observed for an incidence angle of $\theta=10.9\arcdeg$ (see Fig.~\ref{fig4}(a)). For this we use the thickness of the Ni layer as the only free parameter, which we adjust such that the simulation reproduces for each time delay the non-magnetic reflectivity $I_{NM}$.
The feasibility of this procedure is demonstrated by the agreement shown in Fig.~\ref{fig4}(a) between the light green solid line connecting the simulated points and the experimental data (solid symbols). The evolution of the Ni layer thickness underlying this simulation is shown in Fig.~\ref{fig4}(b) and is composed of two phases. An initial compression of up to about 1$\%$ accompanies the ultrafast quenching of the magnetization. This is followed by a slower expansion of up to about 2$\%$ occurring on the subsequent picosecond time scale.

Using  the derived temporal evolution of the Ni layer thickness as input parameter we can calculate how the non-magnetic reflectivity should evolve in the case of the other three grazing incidence angles. Fig.~\ref{fig4}(a) shows that these calculations are overall in good agreement with the experimental data. In particular, the calculation reproduces for all three incidence angles (4.8$\arcdeg$, 6.0$\arcdeg$ and 7.0$\arcdeg$) the observed sign of the reflectivity change for longer time delays.

Regarding the reflectivity curve in Fig.~\ref{fig2}(b) we can actually understand why a variation of the Ni layer thickness causes a strong change in reflectivity only for an incidence angle of $\theta=10.9\arcdeg$. The particularity of this angle is that the reflectivity changes drastically around this angle, which implies that also small thickness variations give rise to significant reflectivity changes. The other three employed incidence angles, on the other hand, fall rather close to the apex of an oscillation and the reflectivity is thus less sensitive to small thickness variations.

In the above analysis we neglected any potential variation of the thickness of the Pd cap and buffer layer. In case of the buffer layer this approximation is justified, since the lower interface of the thick Pd buffer layer does not contribute significantly to the sample's X-ray reflectivity at the employed grazing incidence angles. The cap layer, on the other hand, contributes to the overall reflectivity, but since it is rather thin, small changes of its precise thickness will not alter the reflectivity significantly. In line with this expectation, relative thickness variations of up to $3\%$ are needed to reproduce the reflectivity changes in the $\theta=10.9\arcdeg$ grazing incidence data. Applying this evolution of the Pd cap layer thickness for the other three grazing incidence angles, the experimental data recorded  are, however, not reproduced correctly for longer time delays. We therefore conclude that the observed reflectivity changes are dominated by variations of the Ni layer thickness, without implying that the thickness of the Pd layer would not change for short delays.

Discussing first the slower, picosecond part of the dynamics, we remark that the proposed film expansion is in agreement with the evolution of the non-magnetic reflectivity observed by La-O-Vorakiat {\em et al.}~\cite{la-o-vorakiat_ultrafast_2012} Although these authors did not perform a quantitative analysis of the observed changes, they could follow the evolution to longer time delays. This enabled them to observe an oscillatory evolution, which time scale matches rather closely the speed of a longitudinal acoustic wave. In view of their result, one may note also in our data (see Fig.~\ref{fig4}(a)) the beginning of an oscillatory behavior with a first maximum around 3.7 ps and a subsequent decrease in reflectivity. We furthermore note that our result is  in agreement with the recent observation by Henighan \textit{et al.},~\cite{henighan_generation_2016} who concluded that the ultrafast demagnetization process is accompanied by a strain wave, which expands with a velocity of a few nm/ps into the film bulk. The time dependence of the non-magnetic reflectivity observed by us thus provides a quantitative measurement of the breathing induced by a longitudinal acoustic strain wave on the picosecond time scale.

In view of the even shorter time scale of the initial reduction in X-ray reflectivity, it seems less evident that the film compression proposed by the model reproduces correctly the laser induced ultrafast dynamics. On this time scale of the first 100 - 200 fs following laser excitation one would expect changes in the non-magnetic X-ray reflectivity to be dominated by modifications of the electronic structure. Indeed, changes of the electronic structure accompanying the ultrafast demagnetization process have been observed before.\cite{stamm_femtosecond_2007, frietsch_disparate_2015} For the case of a thin Ni film, Stamm {\em et al.}~\cite{stamm_femtosecond_2007} have observed in their XMCD experiment realized at the BESSY femtoslicing source a shift of the absorption edge to smaller values by about 150 meV. Later, similar shifts have been observed for other magnetic films,\cite{carva_influence_2009, boeglin_distinguishing_2010} too. We note, however, that these shifts are negligible in comparison to the photon energy bandwidth of $\Delta E= 4$ eV used in our experiment. It is therefore not surprising that we cannot model these fast X-ray reflectivity changes with a shifted photon energy scale alone. The same holds true when modeling with significant larger variations of real and imaginary part of the optical constant. A more thorough theoretical analysis of this experimental observation is thus needed. 

\subsection{Challenging predictions of demagnetization models}
The accurate description of our sample structure enables us to model how the X-ray reflectivity should evolve based on the mechanism assumed to drive the ultrafast demagnetization process. One of the most broadly accepted models is based on Elliott-Yafet like splin-flip scattering.\cite{koopmans_explaining_2010, gunther_testing_2014} More recently, it was proposed that the rapid magnetization quenching were due to superdiffusive spin transport out of the excited/probed sample area by the excited spin polarized valence electrons.\cite{battiato_superdiffusive_2010} Several experiments gave evidence for the presence of this mechanism.\cite{malinowski_control_2008, vodungbo_laser-induced_2012, pfau_ultrafast_2012, rudolf_ultrafast_2012, graves_nanoscale_2013} Others propose a co-existence of both mechanisms,\cite{wieczorek_separation_2015, elyasi_optically_2016} but no quantitative assessment of their relative importance/contribution has been obtained so far. 

The evolution of the magnetization profile perpendicular to the film surface predicted by these two models are distinctly different: superdiffusive transport is a non-local phenomenon and a strong demagnetization in the vicinity of the buried interfaces is predicted, which subsequently propagates into the film bulk.\cite{battiato_superdiffusive_2010, battiato_theory_2012, eschenlohr_ultrafast_2013} Elliott-Yafet scattering on the other hand gives rise to a local reduction of the magnetization and the evolution of the magnetization depth profile is determined by the profile of the locally absorbed IR energy.\cite{wieczorek_separation_2015} We note that hot electron-electron collision will wash out this excitation profile, and the evolution of the depth profile will depend on the ratio of the time scales of these two processes. We thus include for our simulation a third model assuming a homogeneous diminution of the magnetization.

\begin{center}
\begin{figure}
 \includegraphics[width=0.48\textwidth]{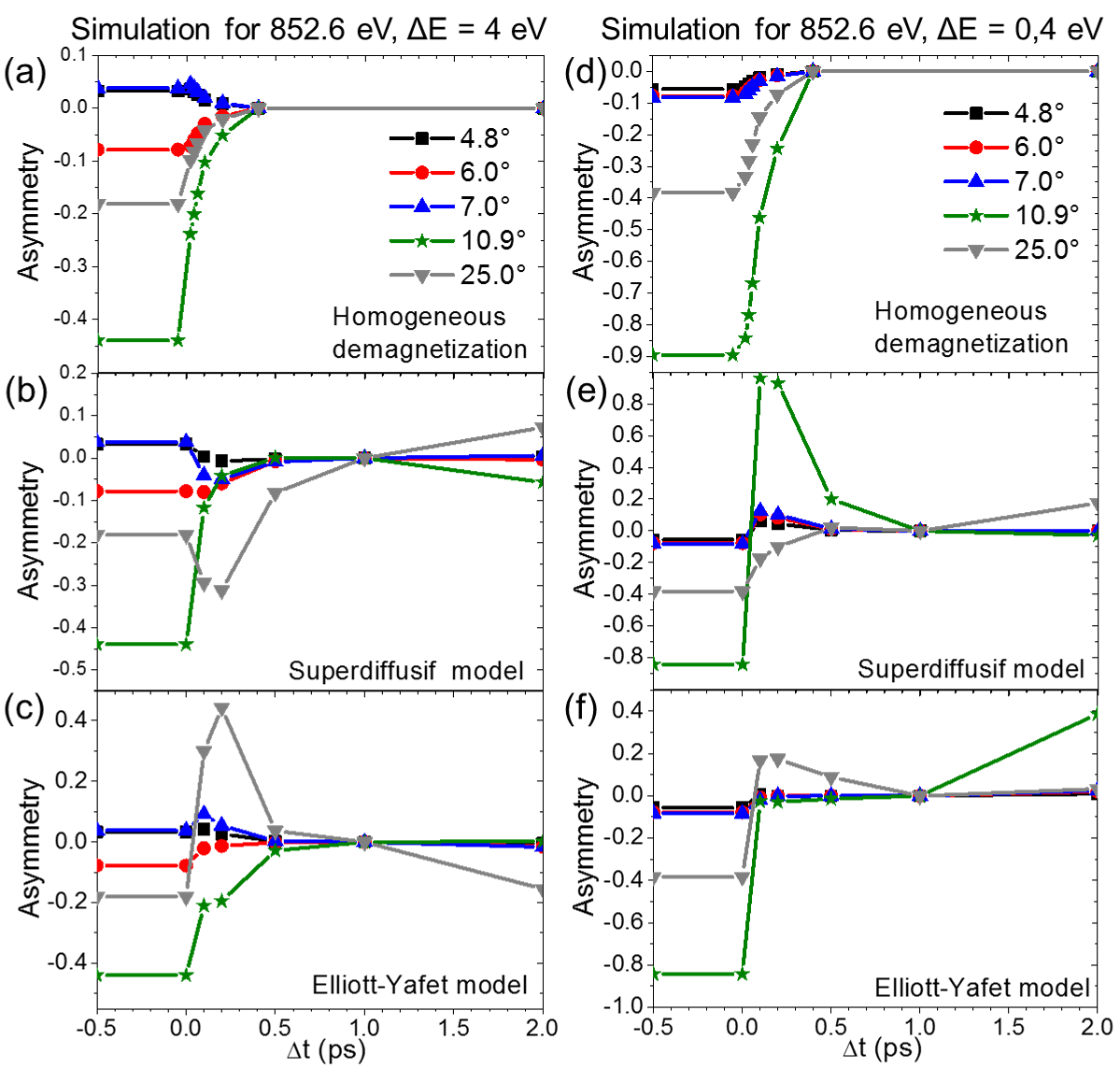}%
 \caption{Evolution of the asymmetry time traces as expected for a model based on a homogeneous demagnetization ((a) (d)), superdiffusive ((b), (e)) and Elliott-Yafet scattering driven demagnetization ((c), (f)). Simulations (a), (b), (c) were performed with an X-ray photon energy bandwidth of $\Delta E = 4$ eV, while a higher photon energy resolution of $\Delta E = 0.4$ eV was used for (d), (e), (f). In each panel the asymmetry time traces are plotted for the four measured incidence angles of $\theta=4.8\arcdeg$ (black square), $\theta=6.0\arcdeg$ (red points), $\theta=7.0\arcdeg$ (blue triangles) and $\theta=10.9\arcdeg$ (green stars) as well as a larger incidence angle of $\theta=25.0\arcdeg$ (gray inverse triangles). 
 \label{fig5}}
 \end{figure}
\end{center}

Using the structural model derived in section C, we can simulate for each of these three models how the magnetic asymmetry is expected to evolve with pump-probe delay. To reproduce an inhomogeneous demagnetization within the Ni layer, we represent it by a series of sublayers each of which a magnetization value is attributed as predicted by the respective demagnetization model.
In Fig.~\ref{fig5} the colored symbols show the result of these calculations for the four probed incidence angles. For the simulations shown in the left column (panel (a), (b), (c)) a photon energy bandwidth of $\Delta E = 4$ eV is assumed. One notices that the asymmetry curves predicted by these three models are overall very similar. And we note that within the present signal-to-noise ratio, they agree all with our experimental data (see Fig.~\ref{fig3}(a)). This implies that based on the current data it is not possible to distinguish between the different demagnetization models.

Sufficient sensitivity to distinguish between these models experimentally is, however, obtained when extending the simulations to larger incidence angles, which corresponds to higher spatial resolution. This is demonstrated by the calculation performed for an incidence angle of $\theta = 25.0\arcdeg$, which is shown by the gray triangular symbols in Fig.~\ref{fig5}. We note that such a measurement would require an increase in photon flux with respect to our current experiment, since a higher incidence angle implies a lower reflectivity. Alternatively, one can opt for a lower photon energy bandwidth as indicated by the calculations shown in the panels on the right side of Fig.~\ref{fig5}. In the case of $\Delta$E = $0.4$ eV, significant differences can be found also for the probed incidence angle of $\theta = 10.9\arcdeg$. We note that both, a significant increase in photon flux and a lower photon energy band width can be obtained at X-ray free electron lasers (e.g., the SXR instrument at LCLS \cite{schlotter_soft_2012}).

\subsection{Conclusions }
In conclusion, we showed that time resolved X-ray magnetic reflectivity at femtosecond pulsed X-ray sources is ideally suited to quantitatively probe ultrafast dynamics of magnetic and structural properties within a single experiment. We applied this technique to investigate the laser excited ultrafast demagnetization dynamics in a prototype ferromagnetic thin film of Ni. The data recorded at the BESSY femtoslicing source reveal unambiguously that the ultrafast demagnetization is accompanied by changes in the non-magnetic reflectivity. Modeling the reflectivity data we can reproduce these changes by varying the thickness of the Ni film. Within this model we find that an ultrafast contraction of the film occurs on the very same time scale as the initial rapid magnetization quenching (about first 200 fs). This is followed by a slower expansion of the film, which occurs on the time scale of a few picoseconds. While a few studies reported structural dynamics accompanying the ultrafast demagnetization phenomenon before, we present here the first quantification of their amplitude. Such combined studies will allow to  probe the presence of a link between both degrees of freedom during the ultrafast demagnetization process.

We furthermore demonstrate that time resolved X-ray magnetic reflectivity can be employed to discriminate between different mechanisms proposed to govern ultrafast demagnetization dynamics, since these models differ in their prediction of how the magnetization depth profile shall evolve. We show that the higher photon flux and energy resolution provided by X-ray free electron lasers will yield decisive data to differentiate between the two most broadly discussed models, Elliott-Yafet spin-flip scattering and superdiffusive spin transport.

\bibliography{TR-XRMR_Ni_biblio}

\begin{thebibliography}{48}%
\makeatletter
\providecommand \@ifxundefined [1]{%
 \@ifx{#1\undefined}
}%
\providecommand \@ifnum [1]{%
 \ifnum #1\expandafter \@firstoftwo
 \else \expandafter \@secondoftwo
 \fi
}%
\providecommand \@ifx [1]{%
 \ifx #1\expandafter \@firstoftwo
 \else \expandafter \@secondoftwo
 \fi
}%
\providecommand \natexlab [1]{#1}%
\providecommand \enquote  [1]{``#1''}%
\providecommand \bibnamefont  [1]{#1}%
\providecommand \bibfnamefont [1]{#1}%
\providecommand \citenamefont [1]{#1}%
\providecommand \href@noop [0]{\@secondoftwo}%
\providecommand \href [0]{\begingroup \@sanitize@url \@href}%
\providecommand \@href[1]{\@@startlink{#1}\@@href}%
\providecommand \@@href[1]{\endgroup#1\@@endlink}%
\providecommand \@sanitize@url [0]{\catcode `\\12\catcode `\$12\catcode
  `\&12\catcode `\#12\catcode `\^12\catcode `\_12\catcode `\%12\relax}%
\providecommand \@@startlink[1]{}%
\providecommand \@@endlink[0]{}%
\providecommand \url  [0]{\begingroup\@sanitize@url \@url }%
\providecommand \@url [1]{\endgroup\@href {#1}{\urlprefix }}%
\providecommand \urlprefix  [0]{URL }%
\providecommand \Eprint [0]{\href }%
\providecommand \doibase [0]{http://dx.doi.org/}%
\providecommand \selectlanguage [0]{\@gobble}%
\providecommand \bibinfo  [0]{\@secondoftwo}%
\providecommand \bibfield  [0]{\@secondoftwo}%
\providecommand \translation [1]{[#1]}%
\providecommand \BibitemOpen [0]{}%
\providecommand \bibitemStop [0]{}%
\providecommand \bibitemNoStop [0]{.\EOS\space}%
\providecommand \EOS [0]{\spacefactor3000\relax}%
\providecommand \BibitemShut  [1]{\csname bibitem#1\endcsname}%
\let\auto@bib@innerbib\@empty
\bibitem [{\citenamefont {Beaurepaire}\ \emph {et~al.}(1996)\citenamefont
  {Beaurepaire}, \citenamefont {Merle}, \citenamefont {Daunois},\ and\
  \citenamefont {Bigot}}]{beaurepaire_ultrafast_1996}%
  \BibitemOpen
  \bibfield  {author} {\bibinfo {author} {\bibfnamefont {E.}~\bibnamefont
  {Beaurepaire}}, \bibinfo {author} {\bibfnamefont {J.-C.}\ \bibnamefont
  {Merle}}, \bibinfo {author} {\bibfnamefont {A.}~\bibnamefont {Daunois}}, \
  and\ \bibinfo {author} {\bibfnamefont {J.-Y.}\ \bibnamefont {Bigot}},\ }\href
  {\doibase 10.1103/PhysRevLett.76.4250} {\bibfield  {journal} {\bibinfo
  {journal} {Phys. Rev. Lett.}\ }\textbf {\bibinfo {volume} {76}},\ \bibinfo
  {pages} {4250} (\bibinfo {year} {1996})}\BibitemShut {NoStop}%
\bibitem [{\citenamefont {Stanciu}\ \emph {et~al.}(2007)\citenamefont
  {Stanciu}, \citenamefont {Hansteen}, \citenamefont {Kimel}, \citenamefont
  {Kirilyuk}, \citenamefont {Tsukamoto}, \citenamefont {Itoh},\ and\
  \citenamefont {Rasing}}]{stanciu_all-optical_2007}%
  \BibitemOpen
  \bibfield  {author} {\bibinfo {author} {\bibfnamefont {C.~D.}\ \bibnamefont
  {Stanciu}}, \bibinfo {author} {\bibfnamefont {F.}~\bibnamefont {Hansteen}},
  \bibinfo {author} {\bibfnamefont {A.~V.}\ \bibnamefont {Kimel}}, \bibinfo
  {author} {\bibfnamefont {A.}~\bibnamefont {Kirilyuk}}, \bibinfo {author}
  {\bibfnamefont {A.}~\bibnamefont {Tsukamoto}}, \bibinfo {author}
  {\bibfnamefont {A.}~\bibnamefont {Itoh}}, \ and\ \bibinfo {author}
  {\bibfnamefont {T.}~\bibnamefont {Rasing}},\ }\href {\doibase
  10.1103/PhysRevLett.99.047601} {\bibfield  {journal} {\bibinfo  {journal}
  {Physical Review Letters}\ }\textbf {\bibinfo {volume} {99}},\ \bibinfo
  {pages} {047601} (\bibinfo {year} {2007})}\BibitemShut {NoStop}%
\bibitem [{\citenamefont {Malinowski}\ \emph {et~al.}(2008)\citenamefont
  {Malinowski}, \citenamefont {Dalla~Longa}, \citenamefont {Rietjens},
  \citenamefont {Paluskar}, \citenamefont {Huijink}, \citenamefont {Swagten},\
  and\ \citenamefont {Koopmans}}]{malinowski_control_2008}%
  \BibitemOpen
  \bibfield  {author} {\bibinfo {author} {\bibfnamefont {G.}~\bibnamefont
  {Malinowski}}, \bibinfo {author} {\bibfnamefont {F.}~\bibnamefont
  {Dalla~Longa}}, \bibinfo {author} {\bibfnamefont {J.~H.~H.}\ \bibnamefont
  {Rietjens}}, \bibinfo {author} {\bibfnamefont {P.~V.}\ \bibnamefont
  {Paluskar}}, \bibinfo {author} {\bibfnamefont {R.}~\bibnamefont {Huijink}},
  \bibinfo {author} {\bibfnamefont {H.~J.~M.}\ \bibnamefont {Swagten}}, \ and\
  \bibinfo {author} {\bibfnamefont {B.}~\bibnamefont {Koopmans}},\ }\href
  {\doibase 10.1038/nphys1092} {\bibfield  {journal} {\bibinfo  {journal}
  {Nature Physics}\ }\textbf {\bibinfo {volume} {4}},\ \bibinfo {pages} {855}
  (\bibinfo {year} {2008})}\BibitemShut {NoStop}%
\bibitem [{\citenamefont {Bigot}\ \emph {et~al.}(2009)\citenamefont {Bigot},
  \citenamefont {Vomir},\ and\ \citenamefont
  {Beaurepaire}}]{bigot_coherent_2009}%
  \BibitemOpen
  \bibfield  {author} {\bibinfo {author} {\bibfnamefont {J.-Y.}\ \bibnamefont
  {Bigot}}, \bibinfo {author} {\bibfnamefont {M.}~\bibnamefont {Vomir}}, \ and\
  \bibinfo {author} {\bibfnamefont {E.}~\bibnamefont {Beaurepaire}},\ }\href
  {\doibase 10.1038/nphys1285} {\bibfield  {journal} {\bibinfo  {journal}
  {Nature Physics}\ }\textbf {\bibinfo {volume} {5}},\ \bibinfo {pages} {515}
  (\bibinfo {year} {2009})}\BibitemShut {NoStop}%
\bibitem [{\citenamefont {Koopmans}\ \emph {et~al.}(2010)\citenamefont
  {Koopmans}, \citenamefont {Malinowski}, \citenamefont {Dalla~Longa},
  \citenamefont {Steiauf}, \citenamefont {F\"ahnle}, \citenamefont {Roth},
  \citenamefont {Cinchetti},\ and\ \citenamefont
  {Aeschlimann}}]{koopmans_explaining_2010}%
  \BibitemOpen
  \bibfield  {author} {\bibinfo {author} {\bibfnamefont {B.}~\bibnamefont
  {Koopmans}}, \bibinfo {author} {\bibfnamefont {G.}~\bibnamefont
  {Malinowski}}, \bibinfo {author} {\bibfnamefont {F.}~\bibnamefont
  {Dalla~Longa}}, \bibinfo {author} {\bibfnamefont {D.}~\bibnamefont
  {Steiauf}}, \bibinfo {author} {\bibfnamefont {M.}~\bibnamefont {F\"ahnle}},
  \bibinfo {author} {\bibfnamefont {T.}~\bibnamefont {Roth}}, \bibinfo {author}
  {\bibfnamefont {M.}~\bibnamefont {Cinchetti}}, \ and\ \bibinfo {author}
  {\bibfnamefont {M.}~\bibnamefont {Aeschlimann}},\ }\href {\doibase
  10.1038/nmat2593} {\bibfield  {journal} {\bibinfo  {journal} {Nature
  Materials}\ }\textbf {\bibinfo {volume} {9}},\ \bibinfo {pages} {259}
  (\bibinfo {year} {2010})}\BibitemShut {NoStop}%
\bibitem [{\citenamefont {Battiato}\ \emph {et~al.}(2010)\citenamefont
  {Battiato}, \citenamefont {Carva},\ and\ \citenamefont
  {Oppeneer}}]{battiato_superdiffusive_2010}%
  \BibitemOpen
  \bibfield  {author} {\bibinfo {author} {\bibfnamefont {M.}~\bibnamefont
  {Battiato}}, \bibinfo {author} {\bibfnamefont {K.}~\bibnamefont {Carva}}, \
  and\ \bibinfo {author} {\bibfnamefont {P.~M.}\ \bibnamefont {Oppeneer}},\
  }\href {\doibase 10.1103/PhysRevLett.105.027203} {\bibfield  {journal}
  {\bibinfo  {journal} {Physical Review Letters}\ }\textbf {\bibinfo {volume}
  {105}},\ \bibinfo {pages} {027203} (\bibinfo {year} {2010})}\BibitemShut
  {NoStop}%
\bibitem [{\citenamefont {Boeglin}\ \emph {et~al.}(2010)\citenamefont
  {Boeglin}, \citenamefont {Beaurepaire}, \citenamefont {Halt\'e},
  \citenamefont {L\'opez-Flores}, \citenamefont {Stamm}, \citenamefont
  {Pontius}, \citenamefont {D\"urr},\ and\ \citenamefont
  {Bigot}}]{boeglin_distinguishing_2010}%
  \BibitemOpen
  \bibfield  {author} {\bibinfo {author} {\bibfnamefont {C.}~\bibnamefont
  {Boeglin}}, \bibinfo {author} {\bibfnamefont {E.}~\bibnamefont
  {Beaurepaire}}, \bibinfo {author} {\bibfnamefont {V.}~\bibnamefont
  {Halt\'e}}, \bibinfo {author} {\bibfnamefont {V.}~\bibnamefont
  {L\'opez-Flores}}, \bibinfo {author} {\bibfnamefont {C.}~\bibnamefont
  {Stamm}}, \bibinfo {author} {\bibfnamefont {N.}~\bibnamefont {Pontius}},
  \bibinfo {author} {\bibfnamefont {H.~A.}\ \bibnamefont {D\"urr}}, \ and\
  \bibinfo {author} {\bibfnamefont {J.-Y.}\ \bibnamefont {Bigot}},\ }\href
  {\doibase 10.1038/nature09070} {\bibfield  {journal} {\bibinfo  {journal}
  {Nature}\ }\textbf {\bibinfo {volume} {465}},\ \bibinfo {pages} {458}
  (\bibinfo {year} {2010})}\BibitemShut {NoStop}%
\bibitem [{\citenamefont {Radu}\ \emph {et~al.}(2011)\citenamefont {Radu},
  \citenamefont {Vahaplar}, \citenamefont {Stamm}, \citenamefont {Kachel},
  \citenamefont {Pontius}, \citenamefont {D\"urr}, \citenamefont {Ostler},
  \citenamefont {Barker}, \citenamefont {Evans}, \citenamefont {Chantrell},
  \citenamefont {Tsukamoto}, \citenamefont {Itoh}, \citenamefont {Kirilyuk},
  \citenamefont {Rasing},\ and\ \citenamefont {Kimel}}]{radu_transient_2011}%
  \BibitemOpen
  \bibfield  {author} {\bibinfo {author} {\bibfnamefont {I.}~\bibnamefont
  {Radu}}, \bibinfo {author} {\bibfnamefont {K.}~\bibnamefont {Vahaplar}},
  \bibinfo {author} {\bibfnamefont {C.}~\bibnamefont {Stamm}}, \bibinfo
  {author} {\bibfnamefont {T.}~\bibnamefont {Kachel}}, \bibinfo {author}
  {\bibfnamefont {N.}~\bibnamefont {Pontius}}, \bibinfo {author} {\bibfnamefont
  {H.~A.}\ \bibnamefont {D\"urr}}, \bibinfo {author} {\bibfnamefont {T.~A.}\
  \bibnamefont {Ostler}}, \bibinfo {author} {\bibfnamefont {J.}~\bibnamefont
  {Barker}}, \bibinfo {author} {\bibfnamefont {R.~F.~L.}\ \bibnamefont
  {Evans}}, \bibinfo {author} {\bibfnamefont {R.~W.}\ \bibnamefont
  {Chantrell}}, \bibinfo {author} {\bibfnamefont {A.}~\bibnamefont
  {Tsukamoto}}, \bibinfo {author} {\bibfnamefont {A.}~\bibnamefont {Itoh}},
  \bibinfo {author} {\bibfnamefont {A.}~\bibnamefont {Kirilyuk}}, \bibinfo
  {author} {\bibfnamefont {T.}~\bibnamefont {Rasing}}, \ and\ \bibinfo {author}
  {\bibfnamefont {A.~V.}\ \bibnamefont {Kimel}},\ }\href {\doibase
  10.1038/nature09901} {\bibfield  {journal} {\bibinfo  {journal} {Nature}\
  }\textbf {\bibinfo {volume} {472}},\ \bibinfo {pages} {205} (\bibinfo {year}
  {2011})}\BibitemShut {NoStop}%
\bibitem [{\citenamefont {Graves}\ \emph {et~al.}(2013)\citenamefont {Graves},
  \citenamefont {Reid}, \citenamefont {Wang}, \citenamefont {Wu}, \citenamefont
  {de~Jong}, \citenamefont {Vahaplar}, \citenamefont {Radu}, \citenamefont
  {Bernstein}, \citenamefont {Messerschmidt}, \citenamefont {M\"uller},
  \citenamefont {Coffee}, \citenamefont {Bionta}, \citenamefont {Epp},
  \citenamefont {Hartmann}, \citenamefont {Kimmel}, \citenamefont {Hauser},
  \citenamefont {Hartmann}, \citenamefont {Holl}, \citenamefont {Gorke},
  \citenamefont {Mentink}, \citenamefont {Tsukamoto}, \citenamefont {Fognini},
  \citenamefont {Turner}, \citenamefont {Schlotter}, \citenamefont {Rolles},
  \citenamefont {Soltau}, \citenamefont {Str\"uder}, \citenamefont {Acremann},
  \citenamefont {Kimel}, \citenamefont {Kirilyuk}, \citenamefont {Rasing},
  \citenamefont {St\"ohr}, \citenamefont {Scherz},\ and\ \citenamefont
  {D\"urr}}]{graves_nanoscale_2013}%
  \BibitemOpen
  \bibfield  {author} {\bibinfo {author} {\bibfnamefont {C.~E.}\ \bibnamefont
  {Graves}}, \bibinfo {author} {\bibfnamefont {A.~H.}\ \bibnamefont {Reid}},
  \bibinfo {author} {\bibfnamefont {T.}~\bibnamefont {Wang}}, \bibinfo {author}
  {\bibfnamefont {B.}~\bibnamefont {Wu}}, \bibinfo {author} {\bibfnamefont
  {S.}~\bibnamefont {de~Jong}}, \bibinfo {author} {\bibfnamefont
  {K.}~\bibnamefont {Vahaplar}}, \bibinfo {author} {\bibfnamefont
  {I.}~\bibnamefont {Radu}}, \bibinfo {author} {\bibfnamefont {D.~P.}\
  \bibnamefont {Bernstein}}, \bibinfo {author} {\bibfnamefont {M.}~\bibnamefont
  {Messerschmidt}}, \bibinfo {author} {\bibfnamefont {L.}~\bibnamefont
  {M\"uller}}, \bibinfo {author} {\bibfnamefont {R.}~\bibnamefont {Coffee}},
  \bibinfo {author} {\bibfnamefont {M.}~\bibnamefont {Bionta}}, \bibinfo
  {author} {\bibfnamefont {S.~W.}\ \bibnamefont {Epp}}, \bibinfo {author}
  {\bibfnamefont {R.}~\bibnamefont {Hartmann}}, \bibinfo {author}
  {\bibfnamefont {N.}~\bibnamefont {Kimmel}}, \bibinfo {author} {\bibfnamefont
  {G.}~\bibnamefont {Hauser}}, \bibinfo {author} {\bibfnamefont
  {A.}~\bibnamefont {Hartmann}}, \bibinfo {author} {\bibfnamefont
  {P.}~\bibnamefont {Holl}}, \bibinfo {author} {\bibfnamefont {H.}~\bibnamefont
  {Gorke}}, \bibinfo {author} {\bibfnamefont {J.~H.}\ \bibnamefont {Mentink}},
  \bibinfo {author} {\bibfnamefont {A.}~\bibnamefont {Tsukamoto}}, \bibinfo
  {author} {\bibfnamefont {A.}~\bibnamefont {Fognini}}, \bibinfo {author}
  {\bibfnamefont {J.~J.}\ \bibnamefont {Turner}}, \bibinfo {author}
  {\bibfnamefont {W.~F.}\ \bibnamefont {Schlotter}}, \bibinfo {author}
  {\bibfnamefont {D.}~\bibnamefont {Rolles}}, \bibinfo {author} {\bibfnamefont
  {H.}~\bibnamefont {Soltau}}, \bibinfo {author} {\bibfnamefont
  {L.}~\bibnamefont {Str\"uder}}, \bibinfo {author} {\bibfnamefont
  {Y.}~\bibnamefont {Acremann}}, \bibinfo {author} {\bibfnamefont {A.~V.}\
  \bibnamefont {Kimel}}, \bibinfo {author} {\bibfnamefont {A.}~\bibnamefont
  {Kirilyuk}}, \bibinfo {author} {\bibfnamefont {T.}~\bibnamefont {Rasing}},
  \bibinfo {author} {\bibfnamefont {J.}~\bibnamefont {St\"ohr}}, \bibinfo
  {author} {\bibfnamefont {A.~O.}\ \bibnamefont {Scherz}}, \ and\ \bibinfo
  {author} {\bibfnamefont {H.~A.}\ \bibnamefont {D\"urr}},\ }\href {\doibase
  10.1038/nmat3597} {\bibfield  {journal} {\bibinfo  {journal} {Nature
  Materials}\ }\textbf {\bibinfo {volume} {12}},\ \bibinfo {pages} {293}
  (\bibinfo {year} {2013})}\BibitemShut {NoStop}%
\bibitem [{\citenamefont {Schellekens}\ \emph {et~al.}(2013)\citenamefont
  {Schellekens}, \citenamefont {Verhoeven}, \citenamefont {Vader},\ and\
  \citenamefont {Koopmans}}]{schellekens_investigating_2013}%
  \BibitemOpen
  \bibfield  {author} {\bibinfo {author} {\bibfnamefont {A.~J.}\ \bibnamefont
  {Schellekens}}, \bibinfo {author} {\bibfnamefont {W.}~\bibnamefont
  {Verhoeven}}, \bibinfo {author} {\bibfnamefont {T.~N.}\ \bibnamefont
  {Vader}}, \ and\ \bibinfo {author} {\bibfnamefont {B.}~\bibnamefont
  {Koopmans}},\ }\href {\doibase 10.1063/1.4812658} {\bibfield  {journal}
  {\bibinfo  {journal} {Applied Physics Letters}\ }\textbf {\bibinfo {volume}
  {102}},\ \bibinfo {pages} {252408} (\bibinfo {year} {2013})}\BibitemShut
  {NoStop}%
\bibitem [{\citenamefont {Bergeard}\ \emph {et~al.}(2014)\citenamefont
  {Bergeard}, \citenamefont {L\'opez-Flores}, \citenamefont {Halt\'e},
  \citenamefont {Hehn}, \citenamefont {Stamm}, \citenamefont {Pontius},
  \citenamefont {Beaurepaire},\ and\ \citenamefont
  {Boeglin}}]{bergeard_ultrafast_2014}%
  \BibitemOpen
  \bibfield  {author} {\bibinfo {author} {\bibfnamefont {N.}~\bibnamefont
  {Bergeard}}, \bibinfo {author} {\bibfnamefont {V.}~\bibnamefont
  {L\'opez-Flores}}, \bibinfo {author} {\bibfnamefont {V.}~\bibnamefont
  {Halt\'e}}, \bibinfo {author} {\bibfnamefont {M.}~\bibnamefont {Hehn}},
  \bibinfo {author} {\bibfnamefont {C.}~\bibnamefont {Stamm}}, \bibinfo
  {author} {\bibfnamefont {N.}~\bibnamefont {Pontius}}, \bibinfo {author}
  {\bibfnamefont {E.}~\bibnamefont {Beaurepaire}}, \ and\ \bibinfo {author}
  {\bibfnamefont {C.}~\bibnamefont {Boeglin}},\ }\href {\doibase
  10.1038/ncomms4466} {\bibfield  {journal} {\bibinfo  {journal} {Nature
  Communications}\ }\textbf {\bibinfo {volume} {5}} (\bibinfo {year} {2014}),\
  10.1038/ncomms4466}\BibitemShut {NoStop}%
\bibitem [{\citenamefont {Rose-Petruck}\ \emph {et~al.}(1999)\citenamefont
  {Rose-Petruck}, \citenamefont {Jimenez}, \citenamefont {Guo}, \citenamefont
  {Cavalleri}, \citenamefont {Siders}, \citenamefont {Rksi}, \citenamefont
  {Squier}, \citenamefont {Walker}, \citenamefont {Wilson},\ and\ \citenamefont
  {Barty}}]{rose-petruck_picosecond-milliangstrom_1999}%
  \BibitemOpen
  \bibfield  {author} {\bibinfo {author} {\bibfnamefont {C.}~\bibnamefont
  {Rose-Petruck}}, \bibinfo {author} {\bibfnamefont {R.}~\bibnamefont
  {Jimenez}}, \bibinfo {author} {\bibfnamefont {T.}~\bibnamefont {Guo}},
  \bibinfo {author} {\bibfnamefont {A.}~\bibnamefont {Cavalleri}}, \bibinfo
  {author} {\bibfnamefont {C.~W.}\ \bibnamefont {Siders}}, \bibinfo {author}
  {\bibfnamefont {F.}~\bibnamefont {Rksi}}, \bibinfo {author} {\bibfnamefont
  {J.~A.}\ \bibnamefont {Squier}}, \bibinfo {author} {\bibfnamefont {B.~C.}\
  \bibnamefont {Walker}}, \bibinfo {author} {\bibfnamefont {K.~R.}\
  \bibnamefont {Wilson}}, \ and\ \bibinfo {author} {\bibfnamefont {C.~P.~J.}\
  \bibnamefont {Barty}},\ }\href {\doibase 10.1038/18631} {\bibfield  {journal}
  {\bibinfo  {journal} {Nature}\ }\textbf {\bibinfo {volume} {398}},\ \bibinfo
  {pages} {310} (\bibinfo {year} {1999})}\BibitemShut {NoStop}%
\bibitem [{\citenamefont {Fritz}\ \emph {et~al.}(2007)\citenamefont {Fritz},
  \citenamefont {Reis}, \citenamefont {Adams}, \citenamefont {Akre},
  \citenamefont {Arthur}, \citenamefont {Blome}, \citenamefont {Bucksbaum},
  \citenamefont {Cavalieri}, \citenamefont {Engemann}, \citenamefont {Fahy},
  \citenamefont {Falcone}, \citenamefont {Fuoss}, \citenamefont {Gaffney},
  \citenamefont {George}, \citenamefont {Hajdu}, \citenamefont {Hertlein},
  \citenamefont {Hillyard}, \citenamefont {Horn-von Hoegen}, \citenamefont
  {Kammler}, \citenamefont {Kaspar}, \citenamefont {Kienberger}, \citenamefont
  {Krejcik}, \citenamefont {Lee}, \citenamefont {Lindenberg}, \citenamefont
  {McFarland}, \citenamefont {Meyer}, \citenamefont {Montagne}, \citenamefont
  {Murray}, \citenamefont {Nelson}, \citenamefont {Nicoul}, \citenamefont
  {Pahl}, \citenamefont {Rudati}, \citenamefont {Schlarb}, \citenamefont
  {Siddons}, \citenamefont {Sokolowski-Tinten}, \citenamefont {Tschentscher},
  \citenamefont {von~der Linde},\ and\ \citenamefont
  {Hastings}}]{Fritz_ultrafast_2007}%
  \BibitemOpen
  \bibfield  {author} {\bibinfo {author} {\bibfnamefont {D.~M.}\ \bibnamefont
  {Fritz}}, \bibinfo {author} {\bibfnamefont {D.~A.}\ \bibnamefont {Reis}},
  \bibinfo {author} {\bibfnamefont {B.}~\bibnamefont {Adams}}, \bibinfo
  {author} {\bibfnamefont {R.~A.}\ \bibnamefont {Akre}}, \bibinfo {author}
  {\bibfnamefont {J.}~\bibnamefont {Arthur}}, \bibinfo {author} {\bibfnamefont
  {C.}~\bibnamefont {Blome}}, \bibinfo {author} {\bibfnamefont {P.~H.}\
  \bibnamefont {Bucksbaum}}, \bibinfo {author} {\bibfnamefont {A.~L.}\
  \bibnamefont {Cavalieri}}, \bibinfo {author} {\bibfnamefont {S.}~\bibnamefont
  {Engemann}}, \bibinfo {author} {\bibfnamefont {S.}~\bibnamefont {Fahy}},
  \bibinfo {author} {\bibfnamefont {R.~W.}\ \bibnamefont {Falcone}}, \bibinfo
  {author} {\bibfnamefont {P.~H.}\ \bibnamefont {Fuoss}}, \bibinfo {author}
  {\bibfnamefont {K.~J.}\ \bibnamefont {Gaffney}}, \bibinfo {author}
  {\bibfnamefont {M.~J.}\ \bibnamefont {George}}, \bibinfo {author}
  {\bibfnamefont {J.}~\bibnamefont {Hajdu}}, \bibinfo {author} {\bibfnamefont
  {M.~P.}\ \bibnamefont {Hertlein}}, \bibinfo {author} {\bibfnamefont {P.~B.}\
  \bibnamefont {Hillyard}}, \bibinfo {author} {\bibfnamefont {M.}~\bibnamefont
  {Horn-von Hoegen}}, \bibinfo {author} {\bibfnamefont {M.}~\bibnamefont
  {Kammler}}, \bibinfo {author} {\bibfnamefont {J.}~\bibnamefont {Kaspar}},
  \bibinfo {author} {\bibfnamefont {R.}~\bibnamefont {Kienberger}}, \bibinfo
  {author} {\bibfnamefont {P.}~\bibnamefont {Krejcik}}, \bibinfo {author}
  {\bibfnamefont {S.~H.}\ \bibnamefont {Lee}}, \bibinfo {author} {\bibfnamefont
  {A.~M.}\ \bibnamefont {Lindenberg}}, \bibinfo {author} {\bibfnamefont
  {B.}~\bibnamefont {McFarland}}, \bibinfo {author} {\bibfnamefont
  {D.}~\bibnamefont {Meyer}}, \bibinfo {author} {\bibfnamefont
  {T.}~\bibnamefont {Montagne}}, \bibinfo {author} {\bibfnamefont {{\'E}.~D.}\
  \bibnamefont {Murray}}, \bibinfo {author} {\bibfnamefont {A.~J.}\
  \bibnamefont {Nelson}}, \bibinfo {author} {\bibfnamefont {M.}~\bibnamefont
  {Nicoul}}, \bibinfo {author} {\bibfnamefont {R.}~\bibnamefont {Pahl}},
  \bibinfo {author} {\bibfnamefont {J.}~\bibnamefont {Rudati}}, \bibinfo
  {author} {\bibfnamefont {H.}~\bibnamefont {Schlarb}}, \bibinfo {author}
  {\bibfnamefont {D.~P.}\ \bibnamefont {Siddons}}, \bibinfo {author}
  {\bibfnamefont {K.}~\bibnamefont {Sokolowski-Tinten}}, \bibinfo {author}
  {\bibfnamefont {T.}~\bibnamefont {Tschentscher}}, \bibinfo {author}
  {\bibfnamefont {D.}~\bibnamefont {von~der Linde}}, \ and\ \bibinfo {author}
  {\bibfnamefont {J.~B.}\ \bibnamefont {Hastings}},\ }\href {\doibase
  10.1126/science.1135009} {\bibfield  {journal} {\bibinfo  {journal}
  {Science}\ }\textbf {\bibinfo {volume} {315}},\ \bibinfo {pages} {633}
  (\bibinfo {year} {2007})},\ \Eprint
  {http://arxiv.org/abs/http://science.sciencemag.org/content/315/5812/633.full.pdf}
  {http://science.sciencemag.org/content/315/5812/633.full.pdf} \BibitemShut
  {NoStop}%
\bibitem [{\citenamefont {Korff~Schmising}\ \emph {et~al.}(2008)\citenamefont
  {Korff~Schmising}, \citenamefont {Harpoeth}, \citenamefont {Zhavoronkov},
  \citenamefont {Ansari}, \citenamefont {Aku-Leh}, \citenamefont {Woerner},
  \citenamefont {Elsaesser}, \citenamefont {Bargheer}, \citenamefont
  {Schmidbauer}, \citenamefont {Vrejoiu}, \citenamefont {Hesse},\ and\
  \citenamefont {Alexe}}]{PhysRevB.78.060404}%
  \BibitemOpen
  \bibfield  {author} {\bibinfo {author} {\bibfnamefont {C.~v.}\ \bibnamefont
  {Korff~Schmising}}, \bibinfo {author} {\bibfnamefont {A.}~\bibnamefont
  {Harpoeth}}, \bibinfo {author} {\bibfnamefont {N.}~\bibnamefont
  {Zhavoronkov}}, \bibinfo {author} {\bibfnamefont {Z.}~\bibnamefont {Ansari}},
  \bibinfo {author} {\bibfnamefont {C.}~\bibnamefont {Aku-Leh}}, \bibinfo
  {author} {\bibfnamefont {M.}~\bibnamefont {Woerner}}, \bibinfo {author}
  {\bibfnamefont {T.}~\bibnamefont {Elsaesser}}, \bibinfo {author}
  {\bibfnamefont {M.}~\bibnamefont {Bargheer}}, \bibinfo {author}
  {\bibfnamefont {M.}~\bibnamefont {Schmidbauer}}, \bibinfo {author}
  {\bibfnamefont {I.}~\bibnamefont {Vrejoiu}}, \bibinfo {author} {\bibfnamefont
  {D.}~\bibnamefont {Hesse}}, \ and\ \bibinfo {author} {\bibfnamefont
  {M.}~\bibnamefont {Alexe}},\ }\href {\doibase 10.1103/PhysRevB.78.060404}
  {\bibfield  {journal} {\bibinfo  {journal} {Phys. Rev. B}\ }\textbf {\bibinfo
  {volume} {78}},\ \bibinfo {pages} {060404} (\bibinfo {year}
  {2008})}\BibitemShut {NoStop}%
\bibitem [{\citenamefont {von Reppert}\ \emph {et~al.}(2016)\citenamefont {von
  Reppert}, \citenamefont {Pudell}, \citenamefont {Koc}, \citenamefont
  {Reinhardt}, \citenamefont {Leitenberger}, \citenamefont {Dumesnil},
  \citenamefont {Zamponi},\ and\ \citenamefont
  {Bargheer}}]{von_reppert_persistent_2016}%
  \BibitemOpen
  \bibfield  {author} {\bibinfo {author} {\bibfnamefont {A.}~\bibnamefont {von
  Reppert}}, \bibinfo {author} {\bibfnamefont {J.}~\bibnamefont {Pudell}},
  \bibinfo {author} {\bibfnamefont {A.}~\bibnamefont {Koc}}, \bibinfo {author}
  {\bibfnamefont {M.}~\bibnamefont {Reinhardt}}, \bibinfo {author}
  {\bibfnamefont {W.}~\bibnamefont {Leitenberger}}, \bibinfo {author}
  {\bibfnamefont {K.}~\bibnamefont {Dumesnil}}, \bibinfo {author}
  {\bibfnamefont {F.}~\bibnamefont {Zamponi}}, \ and\ \bibinfo {author}
  {\bibfnamefont {M.}~\bibnamefont {Bargheer}},\ }\href {\doibase
  10.1063/1.4961253} {\bibfield  {journal} {\bibinfo  {journal} {Structural
  Dynamics}\ }\textbf {\bibinfo {volume} {3}},\ \bibinfo {pages} {054302}
  (\bibinfo {year} {2016})}\BibitemShut {NoStop}%
\bibitem [{\citenamefont {Henighan}\ \emph {et~al.}(2016)\citenamefont
  {Henighan}, \citenamefont {Trigo}, \citenamefont {Bonetti}, \citenamefont
  {Granitzka}, \citenamefont {Higley}, \citenamefont {Chen}, \citenamefont
  {Jiang}, \citenamefont {Kukreja}, \citenamefont {Gray}, \citenamefont {Reid},
  \citenamefont {Jal}, \citenamefont {Hoffmann}, \citenamefont {Kozina},
  \citenamefont {Song}, \citenamefont {Chollet}, \citenamefont {Zhu},
  \citenamefont {Xu}, \citenamefont {Jeong}, \citenamefont {Carva},
  \citenamefont {Maldonado}, \citenamefont {Oppeneer}, \citenamefont {Samant},
  \citenamefont {Parkin}, \citenamefont {Reis},\ and\ \citenamefont
  {D\"urr}}]{henighan_generation_2016}%
  \BibitemOpen
  \bibfield  {author} {\bibinfo {author} {\bibfnamefont {T.}~\bibnamefont
  {Henighan}}, \bibinfo {author} {\bibfnamefont {M.}~\bibnamefont {Trigo}},
  \bibinfo {author} {\bibfnamefont {S.}~\bibnamefont {Bonetti}}, \bibinfo
  {author} {\bibfnamefont {P.}~\bibnamefont {Granitzka}}, \bibinfo {author}
  {\bibfnamefont {D.}~\bibnamefont {Higley}}, \bibinfo {author} {\bibfnamefont
  {Z.}~\bibnamefont {Chen}}, \bibinfo {author} {\bibfnamefont {M.~P.}\
  \bibnamefont {Jiang}}, \bibinfo {author} {\bibfnamefont {R.}~\bibnamefont
  {Kukreja}}, \bibinfo {author} {\bibfnamefont {A.}~\bibnamefont {Gray}},
  \bibinfo {author} {\bibfnamefont {A.~H.}\ \bibnamefont {Reid}}, \bibinfo
  {author} {\bibfnamefont {E.}~\bibnamefont {Jal}}, \bibinfo {author}
  {\bibfnamefont {M.~C.}\ \bibnamefont {Hoffmann}}, \bibinfo {author}
  {\bibfnamefont {M.}~\bibnamefont {Kozina}}, \bibinfo {author} {\bibfnamefont
  {S.}~\bibnamefont {Song}}, \bibinfo {author} {\bibfnamefont {M.}~\bibnamefont
  {Chollet}}, \bibinfo {author} {\bibfnamefont {D.}~\bibnamefont {Zhu}},
  \bibinfo {author} {\bibfnamefont {P.~F.}\ \bibnamefont {Xu}}, \bibinfo
  {author} {\bibfnamefont {J.}~\bibnamefont {Jeong}}, \bibinfo {author}
  {\bibfnamefont {K.}~\bibnamefont {Carva}}, \bibinfo {author} {\bibfnamefont
  {P.}~\bibnamefont {Maldonado}}, \bibinfo {author} {\bibfnamefont {P.~M.}\
  \bibnamefont {Oppeneer}}, \bibinfo {author} {\bibfnamefont {M.~G.}\
  \bibnamefont {Samant}}, \bibinfo {author} {\bibfnamefont {S.~S.~P.}\
  \bibnamefont {Parkin}}, \bibinfo {author} {\bibfnamefont {D.~A.}\
  \bibnamefont {Reis}}, \ and\ \bibinfo {author} {\bibfnamefont {H.~A.}\
  \bibnamefont {D\"urr}},\ }\href {\doibase 10.1103/PhysRevB.93.220301}
  {\bibfield  {journal} {\bibinfo  {journal} {Physical Review B}\ }\textbf
  {\bibinfo {volume} {93}},\ \bibinfo {pages} {220301} (\bibinfo {year}
  {2016})}\BibitemShut {NoStop}%
\bibitem [{\citenamefont {Reid}\ \emph {et~al.}(2016)\citenamefont {Reid},
  \citenamefont {Shen}, \citenamefont {Maldonado}, \citenamefont {Chase},
  \citenamefont {Jal}, \citenamefont {Granitzka}, \citenamefont {Carva},
  \citenamefont {Li}, \citenamefont {Li}, \citenamefont {Wu}, \citenamefont
  {Vecchione}, \citenamefont {Liu}, \citenamefont {Chen}, \citenamefont
  {Higley}, \citenamefont {Hartmann}, \citenamefont {Coffee}, \citenamefont
  {Wu}, \citenamefont {Dakowski}, \citenamefont {Schlotter}, \citenamefont
  {Ohldag}, \citenamefont {Takahashi}, \citenamefont {Mehta}, \citenamefont
  {Hellwig}, \citenamefont {Fry}, \citenamefont {Zhu}, \citenamefont {Cao},
  \citenamefont {Fullerton}, \citenamefont {St\"ohr}, \citenamefont {Oppeneer},
  \citenamefont {Wang},\ and\ \citenamefont {D\"urr}}]{reid_ultrafast_2016}%
  \BibitemOpen
  \bibfield  {author} {\bibinfo {author} {\bibfnamefont {A.~H.}\ \bibnamefont
  {Reid}}, \bibinfo {author} {\bibfnamefont {X.}~\bibnamefont {Shen}}, \bibinfo
  {author} {\bibfnamefont {P.}~\bibnamefont {Maldonado}}, \bibinfo {author}
  {\bibfnamefont {T.}~\bibnamefont {Chase}}, \bibinfo {author} {\bibfnamefont
  {E.}~\bibnamefont {Jal}}, \bibinfo {author} {\bibfnamefont {P.}~\bibnamefont
  {Granitzka}}, \bibinfo {author} {\bibfnamefont {K.}~\bibnamefont {Carva}},
  \bibinfo {author} {\bibfnamefont {R.~K.}\ \bibnamefont {Li}}, \bibinfo
  {author} {\bibfnamefont {J.}~\bibnamefont {Li}}, \bibinfo {author}
  {\bibfnamefont {L.}~\bibnamefont {Wu}}, \bibinfo {author} {\bibfnamefont
  {T.}~\bibnamefont {Vecchione}}, \bibinfo {author} {\bibfnamefont
  {T.}~\bibnamefont {Liu}}, \bibinfo {author} {\bibfnamefont {Z.}~\bibnamefont
  {Chen}}, \bibinfo {author} {\bibfnamefont {D.~J.}\ \bibnamefont {Higley}},
  \bibinfo {author} {\bibfnamefont {N.}~\bibnamefont {Hartmann}}, \bibinfo
  {author} {\bibfnamefont {R.}~\bibnamefont {Coffee}}, \bibinfo {author}
  {\bibfnamefont {J.}~\bibnamefont {Wu}}, \bibinfo {author} {\bibfnamefont
  {G.~L.}\ \bibnamefont {Dakowski}}, \bibinfo {author} {\bibfnamefont
  {W.}~\bibnamefont {Schlotter}}, \bibinfo {author} {\bibfnamefont
  {H.}~\bibnamefont {Ohldag}}, \bibinfo {author} {\bibfnamefont {Y.~K.}\
  \bibnamefont {Takahashi}}, \bibinfo {author} {\bibfnamefont {V.}~\bibnamefont
  {Mehta}}, \bibinfo {author} {\bibfnamefont {O.}~\bibnamefont {Hellwig}},
  \bibinfo {author} {\bibfnamefont {A.}~\bibnamefont {Fry}}, \bibinfo {author}
  {\bibfnamefont {Y.}~\bibnamefont {Zhu}}, \bibinfo {author} {\bibfnamefont
  {J.}~\bibnamefont {Cao}}, \bibinfo {author} {\bibfnamefont {E.~E.}\
  \bibnamefont {Fullerton}}, \bibinfo {author} {\bibfnamefont {J.}~\bibnamefont
  {St\"ohr}}, \bibinfo {author} {\bibfnamefont {P.~M.}\ \bibnamefont
  {Oppeneer}}, \bibinfo {author} {\bibfnamefont {X.~J.}\ \bibnamefont {Wang}},
  \ and\ \bibinfo {author} {\bibfnamefont {H.~A.}\ \bibnamefont {D\"urr}},\
  }\href {http://arxiv.org/abs/1602.04519} {\bibfield  {journal} {\bibinfo
  {journal} {arXiv:1602.04519 [cond-mat]}\ } (\bibinfo {year} {2016})},\
  \bibinfo {note} {arXiv: 1602.04519}\BibitemShut {NoStop}%
\bibitem [{\citenamefont {Tonnerre}\ \emph {et~al.}(2011)\citenamefont
  {Tonnerre}, \citenamefont {Przybylski}, \citenamefont {Ragheb}, \citenamefont
  {Yildiz}, \citenamefont {Tolentino}, \citenamefont {Ortega},\ and\
  \citenamefont {Kirschner}}]{tonnerre_direct_2011}%
  \BibitemOpen
  \bibfield  {author} {\bibinfo {author} {\bibfnamefont {J.-M.}\ \bibnamefont
  {Tonnerre}}, \bibinfo {author} {\bibfnamefont {M.}~\bibnamefont
  {Przybylski}}, \bibinfo {author} {\bibfnamefont {M.}~\bibnamefont {Ragheb}},
  \bibinfo {author} {\bibfnamefont {F.}~\bibnamefont {Yildiz}}, \bibinfo
  {author} {\bibfnamefont {H.~C.~N.}\ \bibnamefont {Tolentino}}, \bibinfo
  {author} {\bibfnamefont {L.}~\bibnamefont {Ortega}}, \ and\ \bibinfo {author}
  {\bibfnamefont {J.}~\bibnamefont {Kirschner}},\ }\href {\doibase
  10.1103/PhysRevB.84.100407} {\bibfield  {journal} {\bibinfo  {journal}
  {Physical Review B}\ }\textbf {\bibinfo {volume} {84}},\ \bibinfo {pages}
  {100407} (\bibinfo {year} {2011})}\BibitemShut {NoStop}%
\bibitem [{\citenamefont {Kortright}(2013)}]{kortright_resonant_2013}%
  \BibitemOpen
  \bibfield  {author} {\bibinfo {author} {\bibfnamefont {J.~B.}\ \bibnamefont
  {Kortright}},\ }\href {\doibase 10.1016/j.elspec.2013.01.019} {\bibfield
  {journal} {\bibinfo  {journal} {Journal of Electron Spectroscopy and Related
  Phenomena}\ }\textbf {\bibinfo {volume} {189}},\ \bibinfo {pages} {178}
  (\bibinfo {year} {2013})}\BibitemShut {NoStop}%
\bibitem [{\citenamefont {Jal}\ \emph {et~al.}(2015)\citenamefont {Jal},
  \citenamefont {Dabrowski}, \citenamefont {Tonnerre}, \citenamefont
  {Przybylski}, \citenamefont {Grenier}, \citenamefont {Jaouen},\ and\
  \citenamefont {Kirschner}}]{jal_noncollinearity_2015}%
  \BibitemOpen
  \bibfield  {author} {\bibinfo {author} {\bibfnamefont {E.}~\bibnamefont
  {Jal}}, \bibinfo {author} {\bibfnamefont {M.}~\bibnamefont {Dabrowski}},
  \bibinfo {author} {\bibfnamefont {J.~M.}\ \bibnamefont {Tonnerre}}, \bibinfo
  {author} {\bibfnamefont {M.}~\bibnamefont {Przybylski}}, \bibinfo {author}
  {\bibfnamefont {S.}~\bibnamefont {Grenier}}, \bibinfo {author} {\bibfnamefont
  {N.}~\bibnamefont {Jaouen}}, \ and\ \bibinfo {author} {\bibfnamefont
  {J.}~\bibnamefont {Kirschner}},\ }\href {\doibase 10.1103/PhysRevB.91.214418}
  {\bibfield  {journal} {\bibinfo  {journal} {Physical Review B}\ }\textbf
  {\bibinfo {volume} {91}},\ \bibinfo {pages} {214418} (\bibinfo {year}
  {2015})}\BibitemShut {NoStop}%
\bibitem [{\citenamefont {Jal}\ \emph {et~al.}(2013)\citenamefont {Jal},
  \citenamefont {Dabrowski}, \citenamefont {Tonnerre}, \citenamefont
  {Przybylski}, \citenamefont {Grenier}, \citenamefont {Jaouen},\ and\
  \citenamefont {Kirschner}}]{jal_magnetization_2013}%
  \BibitemOpen
  \bibfield  {author} {\bibinfo {author} {\bibfnamefont {E.}~\bibnamefont
  {Jal}}, \bibinfo {author} {\bibfnamefont {M.}~\bibnamefont {Dabrowski}},
  \bibinfo {author} {\bibfnamefont {J.-M.}\ \bibnamefont {Tonnerre}}, \bibinfo
  {author} {\bibfnamefont {M.}~\bibnamefont {Przybylski}}, \bibinfo {author}
  {\bibfnamefont {S.}~\bibnamefont {Grenier}}, \bibinfo {author} {\bibfnamefont
  {N.}~\bibnamefont {Jaouen}}, \ and\ \bibinfo {author} {\bibfnamefont
  {J.}~\bibnamefont {Kirschner}},\ }\href {\doibase 10.1103/PhysRevB.87.224418}
  {\bibfield  {journal} {\bibinfo  {journal} {Physical Review B}\ }\textbf
  {\bibinfo {volume} {87}},\ \bibinfo {pages} {224418} (\bibinfo {year}
  {2013})}\BibitemShut {NoStop}%
\bibitem [{\citenamefont {Tsuyama}\ \emph {et~al.}(2016)\citenamefont
  {Tsuyama}, \citenamefont {Chakraverty}, \citenamefont {Macke}, \citenamefont
  {Pontius}, \citenamefont {Sch\"u\ss{}ler-Langeheine}, \citenamefont {Hwang},
  \citenamefont {Tokura},\ and\ \citenamefont
  {Wadati}}]{tsuyama_photoinduced_2016}%
  \BibitemOpen
  \bibfield  {author} {\bibinfo {author} {\bibfnamefont {T.}~\bibnamefont
  {Tsuyama}}, \bibinfo {author} {\bibfnamefont {S.}~\bibnamefont
  {Chakraverty}}, \bibinfo {author} {\bibfnamefont {S.}~\bibnamefont {Macke}},
  \bibinfo {author} {\bibfnamefont {N.}~\bibnamefont {Pontius}}, \bibinfo
  {author} {\bibfnamefont {C.}~\bibnamefont {Sch\"u\ss{}ler-Langeheine}},
  \bibinfo {author} {\bibfnamefont {H.~Y.}\ \bibnamefont {Hwang}}, \bibinfo
  {author} {\bibfnamefont {Y.}~\bibnamefont {Tokura}}, \ and\ \bibinfo {author}
  {\bibfnamefont {H.}~\bibnamefont {Wadati}},\ }\href {\doibase
  10.1103/PhysRevLett.116.256402} {\bibfield  {journal} {\bibinfo  {journal}
  {Phys. Rev. Lett.}\ }\textbf {\bibinfo {volume} {116}},\ \bibinfo {pages}
  {256402} (\bibinfo {year} {2016})}\BibitemShut {NoStop}%
\bibitem [{\citenamefont {Idir}\ \emph {et~al.}(2010)\citenamefont {Idir},
  \citenamefont {Mercere}, \citenamefont {Moreno}, \citenamefont {Delmotte},
  \citenamefont {Dasilva}, \citenamefont {Modi}, \citenamefont {Garrett},
  \citenamefont {Gentle}, \citenamefont {Nugent},\ and\ \citenamefont
  {Wilkins}}]{idir_metrology_2010}%
  \BibitemOpen
  \bibfield  {author} {\bibinfo {author} {\bibfnamefont {M.}~\bibnamefont
  {Idir}}, \bibinfo {author} {\bibfnamefont {P.}~\bibnamefont {Mercere}},
  \bibinfo {author} {\bibfnamefont {T.}~\bibnamefont {Moreno}}, \bibinfo
  {author} {\bibfnamefont {A.}~\bibnamefont {Delmotte}}, \bibinfo {author}
  {\bibfnamefont {P.}~\bibnamefont {Dasilva}}, \bibinfo {author} {\bibfnamefont
  {M.~H.}\ \bibnamefont {Modi}}, \bibinfo {author} {\bibfnamefont
  {R.}~\bibnamefont {Garrett}}, \bibinfo {author} {\bibfnamefont
  {I.}~\bibnamefont {Gentle}}, \bibinfo {author} {\bibfnamefont
  {K.}~\bibnamefont {Nugent}}, \ and\ \bibinfo {author} {\bibfnamefont
  {S.}~\bibnamefont {Wilkins}},\ }\href {\doibase 10.1063/1.3463247} {\bibfield
   {journal} {\bibinfo  {journal} {AIP Conference Proceedings}\ }\textbf
  {\bibinfo {volume} {1234}},\ \bibinfo {pages} {485} (\bibinfo {year}
  {2010})}\BibitemShut {NoStop}%
\bibitem [{\citenamefont {Sacchi}\ \emph {et~al.}(2013)\citenamefont {Sacchi},
  \citenamefont {Jaouen}, \citenamefont {Popescu}, \citenamefont {Gaudemer},
  \citenamefont {Tonnerre}, \citenamefont {Chiuzbaian}, \citenamefont {Hague},
  \citenamefont {Delmotte}, \citenamefont {Dubuisson}, \citenamefont {Cauchon},
  \citenamefont {Lagarde},\ and\ \citenamefont
  {Polack}}]{sacchi_sextants_2013}%
  \BibitemOpen
  \bibfield  {author} {\bibinfo {author} {\bibfnamefont {M.}~\bibnamefont
  {Sacchi}}, \bibinfo {author} {\bibfnamefont {N.}~\bibnamefont {Jaouen}},
  \bibinfo {author} {\bibfnamefont {H.}~\bibnamefont {Popescu}}, \bibinfo
  {author} {\bibfnamefont {R.}~\bibnamefont {Gaudemer}}, \bibinfo {author}
  {\bibfnamefont {J.~M.}\ \bibnamefont {Tonnerre}}, \bibinfo {author}
  {\bibfnamefont {S.~G.}\ \bibnamefont {Chiuzbaian}}, \bibinfo {author}
  {\bibfnamefont {C.~F.}\ \bibnamefont {Hague}}, \bibinfo {author}
  {\bibfnamefont {A.}~\bibnamefont {Delmotte}}, \bibinfo {author}
  {\bibfnamefont {J.~M.}\ \bibnamefont {Dubuisson}}, \bibinfo {author}
  {\bibfnamefont {G.}~\bibnamefont {Cauchon}}, \bibinfo {author} {\bibfnamefont
  {B.}~\bibnamefont {Lagarde}}, \ and\ \bibinfo {author} {\bibfnamefont
  {F.}~\bibnamefont {Polack}},\ }\href {\doibase
  10.1088/1742-6596/425/7/072018} {\bibfield  {journal} {\bibinfo  {journal}
  {Journal of Physics: Conference Series}\ }\textbf {\bibinfo {volume} {425}},\
  \bibinfo {pages} {072018} (\bibinfo {year} {2013})}\BibitemShut {NoStop}%
\bibitem [{\citenamefont {Holldack}\ \emph {et~al.}(2014)\citenamefont
  {Holldack}, \citenamefont {Bahrdt}, \citenamefont {Balzer}, \citenamefont
  {Bovensiepen}, \citenamefont {Brzhezinskaya}, \citenamefont {Erko},
  \citenamefont {Eschenlohr}, \citenamefont {Follath}, \citenamefont {Firsov},
  \citenamefont {Frentrup}, \citenamefont {Le~Guyader}, \citenamefont {Kachel},
  \citenamefont {Kuske}, \citenamefont {Mitzner}, \citenamefont {M\"uller},
  \citenamefont {Pontius}, \citenamefont {Quast}, \citenamefont {Radu},
  \citenamefont {Schmidt}, \citenamefont {Sch\"u\ss{}ler-Langeheine},
  \citenamefont {Sperling}, \citenamefont {Stamm}, \citenamefont {Trabant},\
  and\ \citenamefont {F\"ohlisch}}]{holldack_femtospex:_2014}%
  \BibitemOpen
  \bibfield  {author} {\bibinfo {author} {\bibfnamefont {K.}~\bibnamefont
  {Holldack}}, \bibinfo {author} {\bibfnamefont {J.}~\bibnamefont {Bahrdt}},
  \bibinfo {author} {\bibfnamefont {A.}~\bibnamefont {Balzer}}, \bibinfo
  {author} {\bibfnamefont {U.}~\bibnamefont {Bovensiepen}}, \bibinfo {author}
  {\bibfnamefont {M.}~\bibnamefont {Brzhezinskaya}}, \bibinfo {author}
  {\bibfnamefont {A.}~\bibnamefont {Erko}}, \bibinfo {author} {\bibfnamefont
  {A.}~\bibnamefont {Eschenlohr}}, \bibinfo {author} {\bibfnamefont
  {R.}~\bibnamefont {Follath}}, \bibinfo {author} {\bibfnamefont
  {A.}~\bibnamefont {Firsov}}, \bibinfo {author} {\bibfnamefont
  {W.}~\bibnamefont {Frentrup}}, \bibinfo {author} {\bibfnamefont
  {L.}~\bibnamefont {Le~Guyader}}, \bibinfo {author} {\bibfnamefont
  {T.}~\bibnamefont {Kachel}}, \bibinfo {author} {\bibfnamefont
  {P.}~\bibnamefont {Kuske}}, \bibinfo {author} {\bibfnamefont
  {R.}~\bibnamefont {Mitzner}}, \bibinfo {author} {\bibfnamefont
  {R.}~\bibnamefont {M\"uller}}, \bibinfo {author} {\bibfnamefont
  {N.}~\bibnamefont {Pontius}}, \bibinfo {author} {\bibfnamefont
  {T.}~\bibnamefont {Quast}}, \bibinfo {author} {\bibfnamefont
  {I.}~\bibnamefont {Radu}}, \bibinfo {author} {\bibfnamefont {J.-S.}\
  \bibnamefont {Schmidt}}, \bibinfo {author} {\bibfnamefont {C.}~\bibnamefont
  {Sch\"u\ss{}ler-Langeheine}}, \bibinfo {author} {\bibfnamefont
  {M.}~\bibnamefont {Sperling}}, \bibinfo {author} {\bibfnamefont
  {C.}~\bibnamefont {Stamm}}, \bibinfo {author} {\bibfnamefont
  {C.}~\bibnamefont {Trabant}}, \ and\ \bibinfo {author} {\bibfnamefont
  {A.}~\bibnamefont {F\"ohlisch}},\ }\href {\doibase 10.1107/S1600577514012247}
  {\bibfield  {journal} {\bibinfo  {journal} {Journal of Synchrotron
  Radiation}\ }\textbf {\bibinfo {volume} {21}},\ \bibinfo {pages} {1090}
  (\bibinfo {year} {2014})}\BibitemShut {NoStop}%
\bibitem [{\citenamefont {Schick}\ \emph {et~al.}(2016)\citenamefont {Schick},
  \citenamefont {Le~Guyader}, \citenamefont {Pontius}, \citenamefont {Radu},
  \citenamefont {Kachel}, \citenamefont {Mitzner}, \citenamefont {Zeschke},
  \citenamefont {Schüßler-Langeheine}, \citenamefont {F\"ohlisch},\ and\
  \citenamefont {Holldack}}]{schick_analysis_2016}%
  \BibitemOpen
  \bibfield  {author} {\bibinfo {author} {\bibfnamefont {D.}~\bibnamefont
  {Schick}}, \bibinfo {author} {\bibfnamefont {L.}~\bibnamefont {Le~Guyader}},
  \bibinfo {author} {\bibfnamefont {N.}~\bibnamefont {Pontius}}, \bibinfo
  {author} {\bibfnamefont {I.}~\bibnamefont {Radu}}, \bibinfo {author}
  {\bibfnamefont {T.}~\bibnamefont {Kachel}}, \bibinfo {author} {\bibfnamefont
  {R.}~\bibnamefont {Mitzner}}, \bibinfo {author} {\bibfnamefont
  {T.}~\bibnamefont {Zeschke}}, \bibinfo {author} {\bibfnamefont
  {C.}~\bibnamefont {Schüßler-Langeheine}}, \bibinfo {author} {\bibfnamefont
  {A.}~\bibnamefont {F\"ohlisch}}, \ and\ \bibinfo {author} {\bibfnamefont
  {K.}~\bibnamefont {Holldack}},\ }\href {\doibase 10.1107/S160057751600401X}
  {\bibfield  {journal} {\bibinfo  {journal} {Journal of Synchrotron
  Radiation}\ }\textbf {\bibinfo {volume} {23}},\ \bibinfo {pages} {700}
  (\bibinfo {year} {2016})}\BibitemShut {NoStop}%
\bibitem [{\citenamefont {Mertins}\ \emph {et~al.}(2002)\citenamefont
  {Mertins}, \citenamefont {Abramsohn}, \citenamefont {Gaupp}, \citenamefont
  {Sch\"afers}, \citenamefont {Gudat}, \citenamefont {Zaharko}, \citenamefont
  {Grimmer},\ and\ \citenamefont {Oppeneer}}]{mertins_resonant_2002}%
  \BibitemOpen
  \bibfield  {author} {\bibinfo {author} {\bibfnamefont {H.-C.}\ \bibnamefont
  {Mertins}}, \bibinfo {author} {\bibfnamefont {D.}~\bibnamefont {Abramsohn}},
  \bibinfo {author} {\bibfnamefont {A.}~\bibnamefont {Gaupp}}, \bibinfo
  {author} {\bibfnamefont {F.}~\bibnamefont {Sch\"afers}}, \bibinfo {author}
  {\bibfnamefont {W.}~\bibnamefont {Gudat}}, \bibinfo {author} {\bibfnamefont
  {O.}~\bibnamefont {Zaharko}}, \bibinfo {author} {\bibfnamefont
  {H.}~\bibnamefont {Grimmer}}, \ and\ \bibinfo {author} {\bibfnamefont
  {P.~M.}\ \bibnamefont {Oppeneer}},\ }\href {\doibase
  10.1103/PhysRevB.66.184404} {\bibfield  {journal} {\bibinfo  {journal} {Phys.
  Rev. B}\ }\textbf {\bibinfo {volume} {66}},\ \bibinfo {pages} {184404}
  (\bibinfo {year} {2002})}\BibitemShut {NoStop}%
\bibitem [{\citenamefont {Tonnerre}\ \emph {et~al.}(2012)\citenamefont
  {Tonnerre}, \citenamefont {Jal}, \citenamefont {Bontempi}, \citenamefont
  {Jaouen}, \citenamefont {Elzo}, \citenamefont {Grenier}, \citenamefont
  {Meyerheim},\ and\ \citenamefont
  {Przybylski}}]{tonnerre_depth-resolved_2012}%
  \BibitemOpen
  \bibfield  {author} {\bibinfo {author} {\bibfnamefont {J.-M.}\ \bibnamefont
  {Tonnerre}}, \bibinfo {author} {\bibfnamefont {E.}~\bibnamefont {Jal}},
  \bibinfo {author} {\bibfnamefont {E.}~\bibnamefont {Bontempi}}, \bibinfo
  {author} {\bibfnamefont {N.}~\bibnamefont {Jaouen}}, \bibinfo {author}
  {\bibfnamefont {M.}~\bibnamefont {Elzo}}, \bibinfo {author} {\bibfnamefont
  {S.}~\bibnamefont {Grenier}}, \bibinfo {author} {\bibfnamefont {H.~L.}\
  \bibnamefont {Meyerheim}}, \ and\ \bibinfo {author} {\bibfnamefont
  {M.}~\bibnamefont {Przybylski}},\ }\href {\doibase
  10.1140/epjst/e2012-01618-y} {\bibfield  {journal} {\bibinfo  {journal} {Eur.
  Phys. J. Spec. Top.}\ }\textbf {\bibinfo {volume} {208}},\ \bibinfo {pages}
  {177} (\bibinfo {year} {2012})}\BibitemShut {NoStop}%
\bibitem [{\citenamefont {As}()}]{note}%
  \BibitemOpen
  \bibfield  {author} {\bibinfo {author} {\bibnamefont {As}},\ }\href@noop {}
  {}\bibinfo {note} {Shown by Kortright \cite{kortright_resonant_2013} the
  resonant magnetic x-ray reflectivity is proportional to $C^2 \pm C \cdot S$
  where $C$ indicates the pure charge and $S$ the pure magnetic contribution.
  Note that $S^2$ is generally negligible and thus omitted. The non-magnetic
  contribution $I_{NM} \doteq I_{ave}=\frac{I_p + I_n}{2}$ is therefore
  proportional to $C^2$ and the magnetic asymmetry $A= \frac{I_p - I_n}{I_p +
  I_n}$ is proportional to $S/C$. The pure magnetic contribution is therefore
  given as $S = A \cdot \sqrt{I_{NM}}$.}\BibitemShut {Stop}%
\bibitem [{\citenamefont {Nakajima}\ \emph {et~al.}(1999)\citenamefont
  {Nakajima}, \citenamefont {St\"ohr},\ and\ \citenamefont
  {Idzerda}}]{nakajima_electron-yield_1999}%
  \BibitemOpen
  \bibfield  {author} {\bibinfo {author} {\bibfnamefont {R.}~\bibnamefont
  {Nakajima}}, \bibinfo {author} {\bibfnamefont {J.}~\bibnamefont {St\"ohr}}, \
  and\ \bibinfo {author} {\bibfnamefont {Y.~U.}\ \bibnamefont {Idzerda}},\
  }\href {\doibase 10.1103/PhysRevB.59.6421} {\bibfield  {journal} {\bibinfo
  {journal} {Physical Review B}\ }\textbf {\bibinfo {volume} {59}},\ \bibinfo
  {pages} {6421} (\bibinfo {year} {1999})}\BibitemShut {NoStop}%
\bibitem [{\citenamefont {Stamm}\ \emph {et~al.}(2007)\citenamefont {Stamm},
  \citenamefont {Kachel}, \citenamefont {Pontius}, \citenamefont {Mitzner},
  \citenamefont {Quast}, \citenamefont {Holldack}, \citenamefont {Khan},
  \citenamefont {Lupulescu}, \citenamefont {Aziz}, \citenamefont {Wietstruk},
  \citenamefont {D\"urr},\ and\ \citenamefont
  {Eberhardt}}]{stamm_femtosecond_2007}%
  \BibitemOpen
  \bibfield  {author} {\bibinfo {author} {\bibfnamefont {C.}~\bibnamefont
  {Stamm}}, \bibinfo {author} {\bibfnamefont {T.}~\bibnamefont {Kachel}},
  \bibinfo {author} {\bibfnamefont {N.}~\bibnamefont {Pontius}}, \bibinfo
  {author} {\bibfnamefont {R.}~\bibnamefont {Mitzner}}, \bibinfo {author}
  {\bibfnamefont {T.}~\bibnamefont {Quast}}, \bibinfo {author} {\bibfnamefont
  {K.}~\bibnamefont {Holldack}}, \bibinfo {author} {\bibfnamefont
  {S.}~\bibnamefont {Khan}}, \bibinfo {author} {\bibfnamefont {C.}~\bibnamefont
  {Lupulescu}}, \bibinfo {author} {\bibfnamefont {E.~F.}\ \bibnamefont {Aziz}},
  \bibinfo {author} {\bibfnamefont {M.}~\bibnamefont {Wietstruk}}, \bibinfo
  {author} {\bibfnamefont {H.~A.}\ \bibnamefont {D\"urr}}, \ and\ \bibinfo
  {author} {\bibfnamefont {W.}~\bibnamefont {Eberhardt}},\ }\href {\doibase
  10.1038/nmat1985} {\bibfield  {journal} {\bibinfo  {journal} {Nature
  Materials}\ }\textbf {\bibinfo {volume} {6}},\ \bibinfo {pages} {740}
  (\bibinfo {year} {2007})}\BibitemShut {NoStop}%
\bibitem [{\citenamefont {Frietsch}\ \emph {et~al.}(2015)\citenamefont
  {Frietsch}, \citenamefont {Bowlan}, \citenamefont {Carley}, \citenamefont
  {Teichmann}, \citenamefont {Wienholdt}, \citenamefont {Hinzke}, \citenamefont
  {Nowak}, \citenamefont {Carva}, \citenamefont {Oppeneer},\ and\ \citenamefont
  {Weinelt}}]{frietsch_disparate_2015}%
  \BibitemOpen
  \bibfield  {author} {\bibinfo {author} {\bibfnamefont {B.}~\bibnamefont
  {Frietsch}}, \bibinfo {author} {\bibfnamefont {J.}~\bibnamefont {Bowlan}},
  \bibinfo {author} {\bibfnamefont {R.}~\bibnamefont {Carley}}, \bibinfo
  {author} {\bibfnamefont {M.}~\bibnamefont {Teichmann}}, \bibinfo {author}
  {\bibfnamefont {S.}~\bibnamefont {Wienholdt}}, \bibinfo {author}
  {\bibfnamefont {D.}~\bibnamefont {Hinzke}}, \bibinfo {author} {\bibfnamefont
  {U.}~\bibnamefont {Nowak}}, \bibinfo {author} {\bibfnamefont
  {K.}~\bibnamefont {Carva}}, \bibinfo {author} {\bibfnamefont {P.~M.}\
  \bibnamefont {Oppeneer}}, \ and\ \bibinfo {author} {\bibfnamefont
  {M.}~\bibnamefont {Weinelt}},\ }\href {\doibase 10.1038/ncomms9262}
  {\bibfield  {journal} {\bibinfo  {journal} {Nature Communications}\ }\textbf
  {\bibinfo {volume} {6}},\ \bibinfo {pages} {8262} (\bibinfo {year}
  {2015})}\BibitemShut {NoStop}%
\bibitem [{\citenamefont {Elzo}\ \emph {et~al.}(2012)\citenamefont {Elzo},
  \citenamefont {Jal}, \citenamefont {Bunau}, \citenamefont {Grenier},
  \citenamefont {Joly}, \citenamefont {Ramos}, \citenamefont {Tolentino},
  \citenamefont {Tonnerre},\ and\ \citenamefont {Jaouen}}]{elzo_x-ray_2012}%
  \BibitemOpen
  \bibfield  {author} {\bibinfo {author} {\bibfnamefont {M.}~\bibnamefont
  {Elzo}}, \bibinfo {author} {\bibfnamefont {E.}~\bibnamefont {Jal}}, \bibinfo
  {author} {\bibfnamefont {O.}~\bibnamefont {Bunau}}, \bibinfo {author}
  {\bibfnamefont {S.}~\bibnamefont {Grenier}}, \bibinfo {author} {\bibfnamefont
  {Y.}~\bibnamefont {Joly}}, \bibinfo {author} {\bibfnamefont {A.}~\bibnamefont
  {Ramos}}, \bibinfo {author} {\bibfnamefont {H.}~\bibnamefont {Tolentino}},
  \bibinfo {author} {\bibfnamefont {J.}~\bibnamefont {Tonnerre}}, \ and\
  \bibinfo {author} {\bibfnamefont {N.}~\bibnamefont {Jaouen}},\ }\href
  {\doibase 10.1016/j.jmmm.2011.07.019} {\bibfield  {journal} {\bibinfo
  {journal} {Journal of Magnetism and Magnetic Materials}\ }\textbf {\bibinfo
  {volume} {324}},\ \bibinfo {pages} {105} (\bibinfo {year}
  {2012})}\BibitemShut {NoStop}%
\bibitem [{\citenamefont {Jaouen}\ \emph {et~al.}(2004)\citenamefont {Jaouen},
  \citenamefont {Tonnerre}, \citenamefont {Kapoujian}, \citenamefont {Taunier},
  \citenamefont {Roux}, \citenamefont {Raoux},\ and\ \citenamefont
  {Sirotti}}]{jaouen_apparatus_2004}%
  \BibitemOpen
  \bibfield  {author} {\bibinfo {author} {\bibfnamefont {N.}~\bibnamefont
  {Jaouen}}, \bibinfo {author} {\bibfnamefont {J.-M.}\ \bibnamefont
  {Tonnerre}}, \bibinfo {author} {\bibfnamefont {G.}~\bibnamefont {Kapoujian}},
  \bibinfo {author} {\bibfnamefont {P.}~\bibnamefont {Taunier}}, \bibinfo
  {author} {\bibfnamefont {J.-P.}\ \bibnamefont {Roux}}, \bibinfo {author}
  {\bibfnamefont {D.}~\bibnamefont {Raoux}}, \ and\ \bibinfo {author}
  {\bibfnamefont {F.}~\bibnamefont {Sirotti}},\ }\href {\doibase
  10.1107/S0909049504013767} {\bibfield  {journal} {\bibinfo  {journal}
  {Journal of Synchrotron Radiation}\ }\textbf {\bibinfo {volume} {11}},\
  \bibinfo {pages} {353} (\bibinfo {year} {2004})}\BibitemShut {NoStop}%
\bibitem [{\citenamefont {Parratt}(1954)}]{parratt_surface_1954}%
  \BibitemOpen
  \bibfield  {author} {\bibinfo {author} {\bibfnamefont {L.~G.}\ \bibnamefont
  {Parratt}},\ }\href {\doibase 10.1103/PhysRev.95.359} {\bibfield  {journal}
  {\bibinfo  {journal} {Physical Review}\ }\textbf {\bibinfo {volume} {95}},\
  \bibinfo {pages} {359} (\bibinfo {year} {1954})}\BibitemShut {NoStop}%
\bibitem [{nis()}]{nist}%
  \BibitemOpen
  \href {http://physics.nist.gov/PhysRefData/FFast/html/form.html} {\enquote
  {\bibinfo {title} {{NIST} {X}-{Ray} {Form} {Factor}, {Atten}. {Scatt}.
  {Tables} {Form} {Page}},}\ }\BibitemShut {NoStop}%
\bibitem [{\citenamefont {Chen}\ \emph {et~al.}(1990)\citenamefont {Chen},
  \citenamefont {Sette}, \citenamefont {Ma},\ and\ \citenamefont
  {Modesti}}]{chen_soft-x-ray_1990}%
  \BibitemOpen
  \bibfield  {author} {\bibinfo {author} {\bibfnamefont {C.~T.}\ \bibnamefont
  {Chen}}, \bibinfo {author} {\bibfnamefont {F.}~\bibnamefont {Sette}},
  \bibinfo {author} {\bibfnamefont {Y.}~\bibnamefont {Ma}}, \ and\ \bibinfo
  {author} {\bibfnamefont {S.}~\bibnamefont {Modesti}},\ }\href {\doibase
  10.1103/PhysRevB.42.7262} {\bibfield  {journal} {\bibinfo  {journal}
  {Physical Review B}\ }\textbf {\bibinfo {volume} {42}},\ \bibinfo {pages}
  {7262} (\bibinfo {year} {1990})}\BibitemShut {NoStop}%
\bibitem [{\citenamefont {La-O-Vorakiat}\ \emph {et~al.}(2012)\citenamefont
  {La-O-Vorakiat}, \citenamefont {Turgut}, \citenamefont {Teale}, \citenamefont
  {Kapteyn}, \citenamefont {Murnane}, \citenamefont {Mathias}, \citenamefont
  {Aeschlimann}, \citenamefont {Schneider}, \citenamefont {Shaw}, \citenamefont
  {Nembach},\ and\ \citenamefont {Silva}}]{la-o-vorakiat_ultrafast_2012}%
  \BibitemOpen
  \bibfield  {author} {\bibinfo {author} {\bibfnamefont {C.}~\bibnamefont
  {La-O-Vorakiat}}, \bibinfo {author} {\bibfnamefont {E.}~\bibnamefont
  {Turgut}}, \bibinfo {author} {\bibfnamefont {C.~A.}\ \bibnamefont {Teale}},
  \bibinfo {author} {\bibfnamefont {H.~C.}\ \bibnamefont {Kapteyn}}, \bibinfo
  {author} {\bibfnamefont {M.~M.}\ \bibnamefont {Murnane}}, \bibinfo {author}
  {\bibfnamefont {S.}~\bibnamefont {Mathias}}, \bibinfo {author} {\bibfnamefont
  {M.}~\bibnamefont {Aeschlimann}}, \bibinfo {author} {\bibfnamefont {C.~M.}\
  \bibnamefont {Schneider}}, \bibinfo {author} {\bibfnamefont {J.~M.}\
  \bibnamefont {Shaw}}, \bibinfo {author} {\bibfnamefont {H.~T.}\ \bibnamefont
  {Nembach}}, \ and\ \bibinfo {author} {\bibfnamefont {T.~J.}\ \bibnamefont
  {Silva}},\ }\href {\doibase 10.1103/PhysRevX.2.011005} {\bibfield  {journal}
  {\bibinfo  {journal} {Physical Review X}\ }\textbf {\bibinfo {volume} {2}},\
  \bibinfo {pages} {011005} (\bibinfo {year} {2012})}\BibitemShut {NoStop}%
\bibitem [{\citenamefont {Carva}\ \emph {et~al.}(2009)\citenamefont {Carva},
  \citenamefont {Legut},\ and\ \citenamefont
  {Oppeneer}}]{carva_influence_2009}%
  \BibitemOpen
  \bibfield  {author} {\bibinfo {author} {\bibfnamefont {K.}~\bibnamefont
  {Carva}}, \bibinfo {author} {\bibfnamefont {D.}~\bibnamefont {Legut}}, \ and\
  \bibinfo {author} {\bibfnamefont {P.~M.}\ \bibnamefont {Oppeneer}},\ }\href
  {\doibase 10.1209/0295-5075/86/57002} {\bibfield  {journal} {\bibinfo
  {journal} {EPL (Europhysics Letters)}\ }\textbf {\bibinfo {volume} {86}},\
  \bibinfo {pages} {57002} (\bibinfo {year} {2009})}\BibitemShut {NoStop}%
\bibitem [{\citenamefont {G\"unther}\ \emph {et~al.}(2014)\citenamefont
  {G\"unther}, \citenamefont {Spezzani}, \citenamefont {Ciprian}, \citenamefont
  {Grazioli}, \citenamefont {Ressel}, \citenamefont {Coreno}, \citenamefont
  {Poletto}, \citenamefont {Miotti}, \citenamefont {Sacchi}, \citenamefont
  {Panaccione}, \citenamefont {Uhlir}, \citenamefont {Fullerton}, \citenamefont
  {De~Ninno},\ and\ \citenamefont {Back}}]{gunther_testing_2014}%
  \BibitemOpen
  \bibfield  {author} {\bibinfo {author} {\bibfnamefont {S.}~\bibnamefont
  {G\"unther}}, \bibinfo {author} {\bibfnamefont {C.}~\bibnamefont {Spezzani}},
  \bibinfo {author} {\bibfnamefont {R.}~\bibnamefont {Ciprian}}, \bibinfo
  {author} {\bibfnamefont {C.}~\bibnamefont {Grazioli}}, \bibinfo {author}
  {\bibfnamefont {B.}~\bibnamefont {Ressel}}, \bibinfo {author} {\bibfnamefont
  {M.}~\bibnamefont {Coreno}}, \bibinfo {author} {\bibfnamefont
  {L.}~\bibnamefont {Poletto}}, \bibinfo {author} {\bibfnamefont
  {P.}~\bibnamefont {Miotti}}, \bibinfo {author} {\bibfnamefont
  {M.}~\bibnamefont {Sacchi}}, \bibinfo {author} {\bibfnamefont
  {G.}~\bibnamefont {Panaccione}}, \bibinfo {author} {\bibfnamefont
  {V.}~\bibnamefont {Uhlir}}, \bibinfo {author} {\bibfnamefont {E.~E.}\
  \bibnamefont {Fullerton}}, \bibinfo {author} {\bibfnamefont {G.}~\bibnamefont
  {De~Ninno}}, \ and\ \bibinfo {author} {\bibfnamefont {C.~H.}\ \bibnamefont
  {Back}},\ }\href {\doibase 10.1103/PhysRevB.90.180407} {\bibfield  {journal}
  {\bibinfo  {journal} {Physical Review B}\ }\textbf {\bibinfo {volume} {90}},\
  \bibinfo {pages} {180407} (\bibinfo {year} {2014})}\BibitemShut {NoStop}%
\bibitem [{\citenamefont {Vodungbo}\ \emph {et~al.}(2012)\citenamefont
  {Vodungbo}, \citenamefont {Gautier}, \citenamefont {Lambert}, \citenamefont
  {Sardinha}, \citenamefont {Lozano}, \citenamefont {Sebban}, \citenamefont
  {Ducousso}, \citenamefont {Boutu}, \citenamefont {Li}, \citenamefont {Tudu},
  \citenamefont {Tortarolo}, \citenamefont {Hawaldar}, \citenamefont
  {Delaunay}, \citenamefont {L\'opez-Flores}, \citenamefont {Arabski},
  \citenamefont {Boeglin}, \citenamefont {Merdji}, \citenamefont {Zeitoun},\
  and\ \citenamefont {L\"uning}}]{vodungbo_laser-induced_2012}%
  \BibitemOpen
  \bibfield  {author} {\bibinfo {author} {\bibfnamefont {B.}~\bibnamefont
  {Vodungbo}}, \bibinfo {author} {\bibfnamefont {J.}~\bibnamefont {Gautier}},
  \bibinfo {author} {\bibfnamefont {G.}~\bibnamefont {Lambert}}, \bibinfo
  {author} {\bibfnamefont {A.~B.}\ \bibnamefont {Sardinha}}, \bibinfo {author}
  {\bibfnamefont {M.}~\bibnamefont {Lozano}}, \bibinfo {author} {\bibfnamefont
  {S.}~\bibnamefont {Sebban}}, \bibinfo {author} {\bibfnamefont
  {M.}~\bibnamefont {Ducousso}}, \bibinfo {author} {\bibfnamefont
  {W.}~\bibnamefont {Boutu}}, \bibinfo {author} {\bibfnamefont
  {K.}~\bibnamefont {Li}}, \bibinfo {author} {\bibfnamefont {B.}~\bibnamefont
  {Tudu}}, \bibinfo {author} {\bibfnamefont {M.}~\bibnamefont {Tortarolo}},
  \bibinfo {author} {\bibfnamefont {R.}~\bibnamefont {Hawaldar}}, \bibinfo
  {author} {\bibfnamefont {R.}~\bibnamefont {Delaunay}}, \bibinfo {author}
  {\bibfnamefont {V.}~\bibnamefont {L\'opez-Flores}}, \bibinfo {author}
  {\bibfnamefont {J.}~\bibnamefont {Arabski}}, \bibinfo {author} {\bibfnamefont
  {C.}~\bibnamefont {Boeglin}}, \bibinfo {author} {\bibfnamefont
  {H.}~\bibnamefont {Merdji}}, \bibinfo {author} {\bibfnamefont
  {P.}~\bibnamefont {Zeitoun}}, \ and\ \bibinfo {author} {\bibfnamefont
  {J.}~\bibnamefont {L\"uning}},\ }\href {\doibase 10.1038/ncomms2007}
  {\bibfield  {journal} {\bibinfo  {journal} {Nature Communications}\ }\textbf
  {\bibinfo {volume} {3}},\ \bibinfo {pages} {999} (\bibinfo {year}
  {2012})}\BibitemShut {NoStop}%
\bibitem [{\citenamefont {Pfau}\ \emph {et~al.}(2012)\citenamefont {Pfau},
  \citenamefont {Schaffert}, \citenamefont {M\"uller}, \citenamefont {Gutt},
  \citenamefont {Al-Shemmary}, \citenamefont {B\"uttner}, \citenamefont
  {Delaunay}, \citenamefont {D\"usterer}, \citenamefont {Flewett},
  \citenamefont {Fr\"omter}, \citenamefont {Geilhufe}, \citenamefont {Guehrs},
  \citenamefont {G\"unther}, \citenamefont {Hawaldar}, \citenamefont {Hille},
  \citenamefont {Jaouen}, \citenamefont {Kobs}, \citenamefont {Li},
  \citenamefont {Mohanty}, \citenamefont {Redlin}, \citenamefont {Schlotter},
  \citenamefont {Stickler}, \citenamefont {Treusch}, \citenamefont {Vodungbo},
  \citenamefont {Kl\"aui}, \citenamefont {Oepen}, \citenamefont {L\"uning},
  \citenamefont {Gr\"ubel},\ and\ \citenamefont
  {Eisebitt}}]{pfau_ultrafast_2012}%
  \BibitemOpen
  \bibfield  {author} {\bibinfo {author} {\bibfnamefont {B.}~\bibnamefont
  {Pfau}}, \bibinfo {author} {\bibfnamefont {S.}~\bibnamefont {Schaffert}},
  \bibinfo {author} {\bibfnamefont {L.}~\bibnamefont {M\"uller}}, \bibinfo
  {author} {\bibfnamefont {C.}~\bibnamefont {Gutt}}, \bibinfo {author}
  {\bibfnamefont {A.}~\bibnamefont {Al-Shemmary}}, \bibinfo {author}
  {\bibfnamefont {F.}~\bibnamefont {B\"uttner}}, \bibinfo {author}
  {\bibfnamefont {R.}~\bibnamefont {Delaunay}}, \bibinfo {author}
  {\bibfnamefont {S.}~\bibnamefont {D\"usterer}}, \bibinfo {author}
  {\bibfnamefont {S.}~\bibnamefont {Flewett}}, \bibinfo {author} {\bibfnamefont
  {R.}~\bibnamefont {Fr\"omter}}, \bibinfo {author} {\bibfnamefont
  {J.}~\bibnamefont {Geilhufe}}, \bibinfo {author} {\bibfnamefont
  {E.}~\bibnamefont {Guehrs}}, \bibinfo {author} {\bibfnamefont {C.~M.}\
  \bibnamefont {G\"unther}}, \bibinfo {author} {\bibfnamefont {R.}~\bibnamefont
  {Hawaldar}}, \bibinfo {author} {\bibfnamefont {M.}~\bibnamefont {Hille}},
  \bibinfo {author} {\bibfnamefont {N.}~\bibnamefont {Jaouen}}, \bibinfo
  {author} {\bibfnamefont {A.}~\bibnamefont {Kobs}}, \bibinfo {author}
  {\bibfnamefont {K.}~\bibnamefont {Li}}, \bibinfo {author} {\bibfnamefont
  {J.}~\bibnamefont {Mohanty}}, \bibinfo {author} {\bibfnamefont
  {H.}~\bibnamefont {Redlin}}, \bibinfo {author} {\bibfnamefont {W.~F.}\
  \bibnamefont {Schlotter}}, \bibinfo {author} {\bibfnamefont {D.}~\bibnamefont
  {Stickler}}, \bibinfo {author} {\bibfnamefont {R.}~\bibnamefont {Treusch}},
  \bibinfo {author} {\bibfnamefont {B.}~\bibnamefont {Vodungbo}}, \bibinfo
  {author} {\bibfnamefont {M.}~\bibnamefont {Kl\"aui}}, \bibinfo {author}
  {\bibfnamefont {H.~P.}\ \bibnamefont {Oepen}}, \bibinfo {author}
  {\bibfnamefont {J.}~\bibnamefont {L\"uning}}, \bibinfo {author}
  {\bibfnamefont {G.}~\bibnamefont {Gr\"ubel}}, \ and\ \bibinfo {author}
  {\bibfnamefont {S.}~\bibnamefont {Eisebitt}},\ }\href {\doibase
  10.1038/ncomms2108} {\bibfield  {journal} {\bibinfo  {journal} {Nature
  Communications}\ }\textbf {\bibinfo {volume} {3}},\ \bibinfo {pages} {1100}
  (\bibinfo {year} {2012})}\BibitemShut {NoStop}%
\bibitem [{\citenamefont {Rudolf}\ \emph {et~al.}(2012)\citenamefont {Rudolf},
  \citenamefont {La-O-Vorakiat}, \citenamefont {Battiato}, \citenamefont
  {Adam}, \citenamefont {Shaw}, \citenamefont {Turgut}, \citenamefont
  {Maldonado}, \citenamefont {Mathias}, \citenamefont {Grychtol}, \citenamefont
  {Nembach}, \citenamefont {Silva}, \citenamefont {Aeschlimann}, \citenamefont
  {Kapteyn}, \citenamefont {Murnane}, \citenamefont {Schneider},\ and\
  \citenamefont {Oppeneer}}]{rudolf_ultrafast_2012}%
  \BibitemOpen
  \bibfield  {author} {\bibinfo {author} {\bibfnamefont {D.}~\bibnamefont
  {Rudolf}}, \bibinfo {author} {\bibfnamefont {C.}~\bibnamefont
  {La-O-Vorakiat}}, \bibinfo {author} {\bibfnamefont {M.}~\bibnamefont
  {Battiato}}, \bibinfo {author} {\bibfnamefont {R.}~\bibnamefont {Adam}},
  \bibinfo {author} {\bibfnamefont {J.~M.}\ \bibnamefont {Shaw}}, \bibinfo
  {author} {\bibfnamefont {E.}~\bibnamefont {Turgut}}, \bibinfo {author}
  {\bibfnamefont {P.}~\bibnamefont {Maldonado}}, \bibinfo {author}
  {\bibfnamefont {S.}~\bibnamefont {Mathias}}, \bibinfo {author} {\bibfnamefont
  {P.}~\bibnamefont {Grychtol}}, \bibinfo {author} {\bibfnamefont {H.~T.}\
  \bibnamefont {Nembach}}, \bibinfo {author} {\bibfnamefont {T.~J.}\
  \bibnamefont {Silva}}, \bibinfo {author} {\bibfnamefont {M.}~\bibnamefont
  {Aeschlimann}}, \bibinfo {author} {\bibfnamefont {H.~C.}\ \bibnamefont
  {Kapteyn}}, \bibinfo {author} {\bibfnamefont {M.~M.}\ \bibnamefont
  {Murnane}}, \bibinfo {author} {\bibfnamefont {C.~M.}\ \bibnamefont
  {Schneider}}, \ and\ \bibinfo {author} {\bibfnamefont {P.~M.}\ \bibnamefont
  {Oppeneer}},\ }\href {\doibase 10.1038/ncomms2029} {\bibfield  {journal}
  {\bibinfo  {journal} {Nature Communications}\ }\textbf {\bibinfo {volume}
  {3}},\ \bibinfo {pages} {1037} (\bibinfo {year} {2012})}\BibitemShut
  {NoStop}%
\bibitem [{\citenamefont {Wieczorek}\ \emph {et~al.}(2015)\citenamefont
  {Wieczorek}, \citenamefont {Eschenlohr}, \citenamefont {Weidtmann},
  \citenamefont {R\"osner}, \citenamefont {Bergeard}, \citenamefont
  {Tarasevitch}, \citenamefont {Wehling},\ and\ \citenamefont
  {Bovensiepen}}]{wieczorek_separation_2015}%
  \BibitemOpen
  \bibfield  {author} {\bibinfo {author} {\bibfnamefont {J.}~\bibnamefont
  {Wieczorek}}, \bibinfo {author} {\bibfnamefont {A.}~\bibnamefont
  {Eschenlohr}}, \bibinfo {author} {\bibfnamefont {B.}~\bibnamefont
  {Weidtmann}}, \bibinfo {author} {\bibfnamefont {M.}~\bibnamefont {R\"osner}},
  \bibinfo {author} {\bibfnamefont {N.}~\bibnamefont {Bergeard}}, \bibinfo
  {author} {\bibfnamefont {A.}~\bibnamefont {Tarasevitch}}, \bibinfo {author}
  {\bibfnamefont {T.~O.}\ \bibnamefont {Wehling}}, \ and\ \bibinfo {author}
  {\bibfnamefont {U.}~\bibnamefont {Bovensiepen}},\ }\href {\doibase
  10.1103/PhysRevB.92.174410} {\bibfield  {journal} {\bibinfo  {journal} {Phys.
  Rev. B}\ }\textbf {\bibinfo {volume} {92}},\ \bibinfo {pages} {174410}
  (\bibinfo {year} {2015})}\BibitemShut {NoStop}%
\bibitem [{\citenamefont {Elyasi}\ and\ \citenamefont
  {Yang}(2016)}]{elyasi_optically_2016}%
  \BibitemOpen
  \bibfield  {author} {\bibinfo {author} {\bibfnamefont {M.}~\bibnamefont
  {Elyasi}}\ and\ \bibinfo {author} {\bibfnamefont {H.}~\bibnamefont {Yang}},\
  }\href {\doibase 10.1103/PhysRevB.94.024417} {\bibfield  {journal} {\bibinfo
  {journal} {Physical Review B}\ }\textbf {\bibinfo {volume} {94}},\ \bibinfo
  {pages} {024417} (\bibinfo {year} {2016})}\BibitemShut {NoStop}%
\bibitem [{\citenamefont {Battiato}\ \emph {et~al.}(2012)\citenamefont
  {Battiato}, \citenamefont {Carva},\ and\ \citenamefont
  {Oppeneer}}]{battiato_theory_2012}%
  \BibitemOpen
  \bibfield  {author} {\bibinfo {author} {\bibfnamefont {M.}~\bibnamefont
  {Battiato}}, \bibinfo {author} {\bibfnamefont {K.}~\bibnamefont {Carva}}, \
  and\ \bibinfo {author} {\bibfnamefont {P.~M.}\ \bibnamefont {Oppeneer}},\
  }\href {\doibase 10.1103/PhysRevB.86.024404} {\bibfield  {journal} {\bibinfo
  {journal} {Physical Review B}\ }\textbf {\bibinfo {volume} {86}},\ \bibinfo
  {pages} {024404} (\bibinfo {year} {2012})}\BibitemShut {NoStop}%
\bibitem [{\citenamefont {Eschenlohr}\ \emph {et~al.}(2013)\citenamefont
  {Eschenlohr}, \citenamefont {Battiato}, \citenamefont {Maldonado},
  \citenamefont {Pontius}, \citenamefont {Kachel}, \citenamefont {Holldack},
  \citenamefont {Mitzner}, \citenamefont {F\"ohlisch}, \citenamefont
  {Oppeneer},\ and\ \citenamefont {Stamm}}]{eschenlohr_ultrafast_2013}%
  \BibitemOpen
  \bibfield  {author} {\bibinfo {author} {\bibfnamefont {A.}~\bibnamefont
  {Eschenlohr}}, \bibinfo {author} {\bibfnamefont {M.}~\bibnamefont
  {Battiato}}, \bibinfo {author} {\bibfnamefont {P.}~\bibnamefont {Maldonado}},
  \bibinfo {author} {\bibfnamefont {N.}~\bibnamefont {Pontius}}, \bibinfo
  {author} {\bibfnamefont {T.}~\bibnamefont {Kachel}}, \bibinfo {author}
  {\bibfnamefont {K.}~\bibnamefont {Holldack}}, \bibinfo {author}
  {\bibfnamefont {R.}~\bibnamefont {Mitzner}}, \bibinfo {author} {\bibfnamefont
  {A.}~\bibnamefont {F\"ohlisch}}, \bibinfo {author} {\bibfnamefont {P.~M.}\
  \bibnamefont {Oppeneer}}, \ and\ \bibinfo {author} {\bibfnamefont
  {C.}~\bibnamefont {Stamm}},\ }\href {\doibase 10.1038/nmat3546} {\bibfield
  {journal} {\bibinfo  {journal} {Nature Materials}\ }\textbf {\bibinfo
  {volume} {12}},\ \bibinfo {pages} {332} (\bibinfo {year} {2013})}\BibitemShut
  {NoStop}%
\bibitem [{\citenamefont {Schlotter}\ \emph {et~al.}(2012)\citenamefont
  {Schlotter}, \citenamefont {Turner}, \citenamefont {Rowen}, \citenamefont
  {Heimann}, \citenamefont {Holmes}, \citenamefont {Krupin}, \citenamefont
  {Messerschmidt}, \citenamefont {Moeller}, \citenamefont {Krzywinski},
  \citenamefont {Soufli}, \citenamefont {Fern\'andez-Perea}, \citenamefont
  {Kelez}, \citenamefont {Lee}, \citenamefont {Coffee}, \citenamefont {Hays},
  \citenamefont {Beye}, \citenamefont {Gerken}, \citenamefont {Sorgenfrei},
  \citenamefont {Hau-Riege}, \citenamefont {Juha}, \citenamefont {Chalupsky},
  \citenamefont {Hajkova}, \citenamefont {Mancuso}, \citenamefont {Singer},
  \citenamefont {Yefanov}, \citenamefont {Vartanyants}, \citenamefont
  {Cadenazzi}, \citenamefont {Abbey}, \citenamefont {Nugent}, \citenamefont
  {Sinn}, \citenamefont {L\"uning}, \citenamefont {Schaffert}, \citenamefont
  {Eisebitt}, \citenamefont {Lee}, \citenamefont {Scherz}, \citenamefont
  {Nilsson},\ and\ \citenamefont {Wurth}}]{schlotter_soft_2012}%
  \BibitemOpen
  \bibfield  {author} {\bibinfo {author} {\bibfnamefont {W.~F.}\ \bibnamefont
  {Schlotter}}, \bibinfo {author} {\bibfnamefont {J.~J.}\ \bibnamefont
  {Turner}}, \bibinfo {author} {\bibfnamefont {M.}~\bibnamefont {Rowen}},
  \bibinfo {author} {\bibfnamefont {P.}~\bibnamefont {Heimann}}, \bibinfo
  {author} {\bibfnamefont {M.}~\bibnamefont {Holmes}}, \bibinfo {author}
  {\bibfnamefont {O.}~\bibnamefont {Krupin}}, \bibinfo {author} {\bibfnamefont
  {M.}~\bibnamefont {Messerschmidt}}, \bibinfo {author} {\bibfnamefont
  {S.}~\bibnamefont {Moeller}}, \bibinfo {author} {\bibfnamefont
  {J.}~\bibnamefont {Krzywinski}}, \bibinfo {author} {\bibfnamefont
  {R.}~\bibnamefont {Soufli}}, \bibinfo {author} {\bibfnamefont
  {M.}~\bibnamefont {Fern\'andez-Perea}}, \bibinfo {author} {\bibfnamefont
  {N.}~\bibnamefont {Kelez}}, \bibinfo {author} {\bibfnamefont
  {S.}~\bibnamefont {Lee}}, \bibinfo {author} {\bibfnamefont {R.}~\bibnamefont
  {Coffee}}, \bibinfo {author} {\bibfnamefont {G.}~\bibnamefont {Hays}},
  \bibinfo {author} {\bibfnamefont {M.}~\bibnamefont {Beye}}, \bibinfo {author}
  {\bibfnamefont {N.}~\bibnamefont {Gerken}}, \bibinfo {author} {\bibfnamefont
  {F.}~\bibnamefont {Sorgenfrei}}, \bibinfo {author} {\bibfnamefont
  {S.}~\bibnamefont {Hau-Riege}}, \bibinfo {author} {\bibfnamefont
  {L.}~\bibnamefont {Juha}}, \bibinfo {author} {\bibfnamefont {J.}~\bibnamefont
  {Chalupsky}}, \bibinfo {author} {\bibfnamefont {V.}~\bibnamefont {Hajkova}},
  \bibinfo {author} {\bibfnamefont {A.~P.}\ \bibnamefont {Mancuso}}, \bibinfo
  {author} {\bibfnamefont {A.}~\bibnamefont {Singer}}, \bibinfo {author}
  {\bibfnamefont {O.}~\bibnamefont {Yefanov}}, \bibinfo {author} {\bibfnamefont
  {I.~A.}\ \bibnamefont {Vartanyants}}, \bibinfo {author} {\bibfnamefont
  {G.}~\bibnamefont {Cadenazzi}}, \bibinfo {author} {\bibfnamefont
  {B.}~\bibnamefont {Abbey}}, \bibinfo {author} {\bibfnamefont {K.~A.}\
  \bibnamefont {Nugent}}, \bibinfo {author} {\bibfnamefont {H.}~\bibnamefont
  {Sinn}}, \bibinfo {author} {\bibfnamefont {J.}~\bibnamefont {L\"uning}},
  \bibinfo {author} {\bibfnamefont {S.}~\bibnamefont {Schaffert}}, \bibinfo
  {author} {\bibfnamefont {S.}~\bibnamefont {Eisebitt}}, \bibinfo {author}
  {\bibfnamefont {W.-S.}\ \bibnamefont {Lee}}, \bibinfo {author} {\bibfnamefont
  {A.}~\bibnamefont {Scherz}}, \bibinfo {author} {\bibfnamefont {A.~R.}\
  \bibnamefont {Nilsson}}, \ and\ \bibinfo {author} {\bibfnamefont
  {W.}~\bibnamefont {Wurth}},\ }\href {\doibase 10.1063/1.3698294} {\bibfield
  {journal} {\bibinfo  {journal} {Review of Scientific Instruments}\ }\textbf
  {\bibinfo {volume} {83}},\ \bibinfo {pages} {043107} (\bibinfo {year}
  {2012})}\BibitemShut {NoStop}%
\end{thebibliography}%

\end{document}